\documentclass[twocolumn,trackchanges]{aastex7}

\usepackage{graphicx}
\usepackage{dcolumn}
\usepackage{bm}
\usepackage{float} %
\usepackage[export]{adjustbox}
\usepackage{amsmath}

\begin{document}

\title{Dark Matter and the Early Formation of Supermassive Black Holes}

\author{Andrew Imai, Grant J. Mathews, Guobao Tang, and Brian Zhang}

\email[show]{aimai@nd.edu}
\email{bzhang8@nd.edu}
\email{gtang@nd.edu}
\email[show]{gmathews@nd.edu}

\affiliation{Center for Astrophysics, Department of Physics and Astronomy, University of Notre Dame, Notre Dame, Indiana, 46556, USA}

\begin{abstract}
We investigate the growth of supermassive black holes (SMBHs) at high redshift ($z \ge 10$)  from a combination of dark matter capture,
 black-hole mergers, and gas accretion. It has previously been shown that SMBHs can form by $z \approx 10$ via black-hole mergers, Eddington-limited Bondi gas accretion and tidal disruption events with stars within dense nuclear clusters. 
Here, we  examine the degree to which the capture of collisionless dark matter by a growing SMBH may also contribute. We first consider models deduced from cosmological simulations of galaxy formation and central BH formation.  We show that in the case that the dense nuclear star cluster forms by cooling and collapse of gas, while the DM remains in a standard NFW profile, the contribution from  cold dark matter accretion  is insignificant.  However, we suggest models  for which dark matter clustering can occur (possibly by self interaction).  We show that such clustering  may affect SMBH growth.   In such cases,  a small seed stellar-remnant black hole can more easily reach $> 10^7$ M$_{\odot}$ by $z = 10$ in the core of dense nuclear star clusters.   This remains true for either cold dark matter or ultralight dark matter with the observationally inferred  mass of $\sim  10^{-22}$ eV.  We highlight the unique possible evolution of ULDM capture by the growing SMBH due to the fact that  the ULDM de Broglie wavelength exceeds the initial nuclear star cluster half-mass radius.

\end{abstract}

\keywords{
\uat{Supermassive black holes}{1663} ---
\uat{Dark matter}{353} ---
\uat{Star clusters}{1567}
}


\section{Introduction} 

A significant problem in modern cosmology is the observation that supermassive black holes (SMBHs) (on the order of $10^6 - 10^8$ M$_{\odot}$) have formed much earlier than expected by stellar black hole formation and accretion \citep{Greene2024, Natarajan24, Juod_balis_2024, Maiolino_2024}. In particular, the James Webb Space Telescope (JWST) has detected a $4 \times 10^7$ M$_\odot$ SMBH in an active galactic nucleus as early as  $z \approx 10$ \citep{Natarajan24}, suggesting that this SMBH could have formed 200-400 Myr after photon decoupling.   In addition, JWST has detected a large number of "Little Red Dots" in the redshift range of $z= 4-8$. These are most likely Active Galactic Nuclei (AGNs) whose energy sources are SMBHs in the range of $10^7-10^8$ M$_\odot$ \citep{Greene2024, Maiolino24, Matthee24}.  All of this indicates that the formation of SMBHs can occur very soon after the big bang.

Various attempts have been made to explain the early formation of SMBHs, including the direct collapse of pristine gas clouds \citep{Agarwal_2014, Latif2022}, accretion rates higher than the Eddington limit (super-Eddington accretion) \citep{Davies_2011, piana2024, trinca2024}, and growth from an initial seed black hole on the order of $10$ M$_{\odot}$ \citep{Begelman2006, Latif2022, Greene2024, kritos2024supe, liempi2024}. In the present work, we expand on the latter scenario by exploring the additional contribution from the capture of Cold Dark Matter (CDM) or ultralight dark matter (ULDM).
 We show that for the most part the inclusion of dark matter makes little difference.  However, we also consider situations in which a dense DM cluster can form along with a dense nuclear star cluster.  We show that  the inclusion of such dark-matter accretion can relax the conditions required for the early formation of supermassive black holes.

\subsection{Nuclear Star Clusters}
We consider the growth of a central seed black hole within a dense and massive nuclear star cluster (NSC) \citep{Biernacki_2017}. Such clusters exist at the center of most galaxies and typically have a mass range of $10^5 - 10^9$ M$_{\odot}$ \citep{Neumayer2020}. Their effective radii are not larger than 2 to 5 parsecs and can be much smaller.  They are a natural environment for the growth of SMBHs \citep{Antonini2012}.

Theoretical studies \citep{Volonteri2008} have suggested that the central regions of proto-galaxies may undergo fragmentation, giving rise to dense stellar clusters that contain millions of metal-poor stars. These clusters are predicted to have very compact structures, with half-mass radii typically less than 1 pc \citep{Devecchi2010}. However,  no clusters with radii smaller than $\sim 1$ parsec have yet been detected \citep{Vanzella2023}.

In this work, we consider a model in which an initial central stellar seed black hole becomes the site for runaway accretion of gas, stars, remnants, and dark matter (DM). This is a variation of the approach of \citep{Kritos2024}.  In that work the combined effects of the black hole mergers and Eddington-limited gas accretion in NSCs were considered to form SMBHs in their numerical code \textsc{NUCE} (Nuclear Cluster Evolution). Here, we modify \textsc{NUCE} to include the presence and evolution of CDM and ULDM. We find that DM capture can dominate overwhelmingly the contribution to the final SMBH mass, while somewhat relaxing the requirement of a high concentration of gas and stars.

 One does not necessarily expect dark matter to be significant in a dense nuclear cluster.  In the standard picture, the formation of gas and  stars in a dense nuclear cluster requires significant radiative cooling and
subsequent contraction of the gas with respect to dark matter.  Hence, the dark matter content is not expected to contribute.  

Nevertheless, there is evidence for high concentrations of dark matter in 'perturbers' such as the $10^6$ M$_\odot$ object that perturbs JVAS B1938+666 through gravitational imaging \citep{Powell25}, or the $10^{5.5}-10^{8}$ M$_\odot$ concentrated mass perturber inferred through observations of the GD-1 stellar stream \citep{Bonaca19,Nibauer25}.  Such objects have been interpreted as core-collapsed self-interacting dark matter halos \citep{Yu26}.  Moreover, these objects can be fit with a truncated isothermal  profile that varies as $\rho(r) \propto r^{-2}$ toward the center.

In this paper, therefore, we speculate that such a core-collapsed SIDM structure may have contributed to the formation and evolution of dense nuclear clusters and the growth of the associated SMBH.  Our goal is not to establish that such structures form, but only to analyze how the possible existence of such a collapsed DM structure could affect SMBH growth.  

\subsection{Initial Conditions}

For the present demonstration, we take the initial seed black hole (BH) to be 30 M$_{\odot}$. This is a conservative estimate of the mass of an initial BH that would form from the collapse of Population III stars in the early galaxy \citep{Xu_2013, Susa_2014}. 
The stars in the model were taken to obey a Kroupa initial mass function \citep{Kroupa02} with a mean stellar mass of 0.59 M$_{\odot}$. 

For simplicity, we consider models that assume a uniform density distribution of DM, stars, and gas within a small volume at the center of the NSC.

 \subsection{Simulation of DM content in NSCs}
 The dark matter distribution in NSCs is not known.  If the dense NSC forms by radiative cooling and collapse of gas, then the dark matter would be unaffected.  However, as noted in the introduction, we also consider the possibility of a collapsed DM structure within the NSC.  Indeed, one could speculate that the formation of such structures may even provide a seed for the formation of the NSC. Hence, we consider the possibility that the DM content could be comparable to or greater than the density of stars, gas, and remnants in the NSC.

As guidance for this, we have identified a $6.1 \times 10^8$ M$_\odot$ SMBH at $z=0$ in the TNG50 run of the IllustrisTNG simulation \citep{Nelson:2019,Pillepich:2019}.    We then analyzed the structure of the pre-SMBH at $z \approx 20$. 

The IllustrisTNG simulations are based upon initial conditions in  a $\Lambda$CDM cosmology with cosmological parameters from the Planck Collaboration \citep{Planck-Collaboration:2016}, i.e.,  $\Omega_m=0.3089$ in which $\Omega_{dm}=0.2603$ and $\Omega_b=0.0486$, $\Omega_\Lambda=0.6911$, $H_0=100 h$ km/s/Mpc with $h=0.6774$, $\sigma_8 = 0.8159$ and $n_s = 0.9667$. The simulations began at $z=127$ and  evolved to the current epoch at $z=0$. The TNG50 simulations use periodic boundary conditions in co-moving coordinates with a box size of 35 Mpc/$h$ (51.7 Mpc).  The smoothed particle hydrodynamics includes a resolution of $8 \times 10^4$ M$_\odot$ for baryonic matter and $4.5 \times 10^5$ M$_\odot$ for DM.  This limits the resolution for small volumes with few particles. Nevertheless, the simulation includes the identification of SMBH formation.

 One should note that such cosmological simulations do not fully resolve the highest central densities. In the TNG50 simulation, the spatial resolution is set by the gravitational softening length (minimum softening scale of $\sim 50$--$70\,\mathrm{pc}$ for gas and $\sim 290\,\mathrm{pc}$ for dark matter and stars at $z=0$; corresponding to a minimum softening scale of $\sim 3$--$4\,\mathrm{pc}$ for gas and $\sim 27\,\mathrm{pc}$ for dark matter and stars at $z \approx 20$). In addition, dense gas above $0.1\,\mathrm{cm}^{-3}$ is modeled using a subgrid multiphase interstellar medium model, which suppresses unresolved collapse on small scales \citep{Pillepich_2017}. While these effects may impact the innermost regions, we use the simulation primarily as a means to plausibly argue that the central DM density could be large. To assess the impact of this suppression, we also consider simulations from the FIRE project \citep{Hopkins_2018}, which allows the gas to cool to lower temperatures and collapse without an imposed effective equation of state.


 In the IllustrisTNG simulations, black holes are seeded at the centers of halos identified by a friends-of-friends (FoF) algorithm once the halo mass exceeds a threshold mass of $5 \times 10^{10}\,h^{-1}\,M_{\odot}$ and no black hole is present \citep{Weinberger_2018}. The seed mass is chosen to be relatively large ($M_{\mathrm{seed}} = 8 \times 10^{5}\,h^{-1}\,M_{\odot}$) to ensure efficient early growth. In the system considered here, the first black hole appears at $z = 6.49$ ($t \approx 840\,\mathrm{Myr}$). 

Once BHs are seeded in massive halos of the TNG50 simulation, they then accrete nearby gas at the Eddington-limited Bondi rate \citep{Nelson:2019}. In the system examined here, a central BH begins to form  during a significant accretion event at $z = 6.49$.  However, in this system, the BH mass does not grow to  $10^7$ M$_\odot$ until about 2 Gyr into the simulation.  Nevertheless, this is a useful example in which to examine the relative contributions of baryonic and dark matter to the growth of the SMBH.

In contrast, the FIRE-2 comparison system used here is drawn from the MassiveFIRE A-series, specifically the A2 zoom-in simulation. This simulation has a baryonic mass resolution of $3.3 \times 10^{4}\,M_{\odot}$ and a dark matter particle mass of $1.7 \times 10^{5}\,M_{\odot}$ \citep{Cochrane2023}. The gravitational softening lengths in FIRE-2 are adaptive for gas, with a minimum physical value of $\sim 0.7\,\mathrm{pc}$ in dense regions, and are fixed at $\sim 7\,\mathrm{pc}$ for stars and $\sim 57\,\mathrm{pc}$ for dark matter \citep{atmabacak_2022}. Black holes are seeded at the location of the most bound star particle once the system reaches $M_* > 1000\,M_{\mathrm{seed}}$, with $M_{\mathrm{seed}} = 1.4 \times 10^{4}\,M_{\odot}$.  Once formed, the SMBHs are evolved using a gravitational torque-limited accretion model \citep{Angl_s_Alc_zar_2017}. In the system considered here, the first black hole appears at $z = 10.4$ ($t \approx 445\,\mathrm{Myr}$).

\begin{figure}[!h]
    \centering
    \includegraphics[width=.9\linewidth]{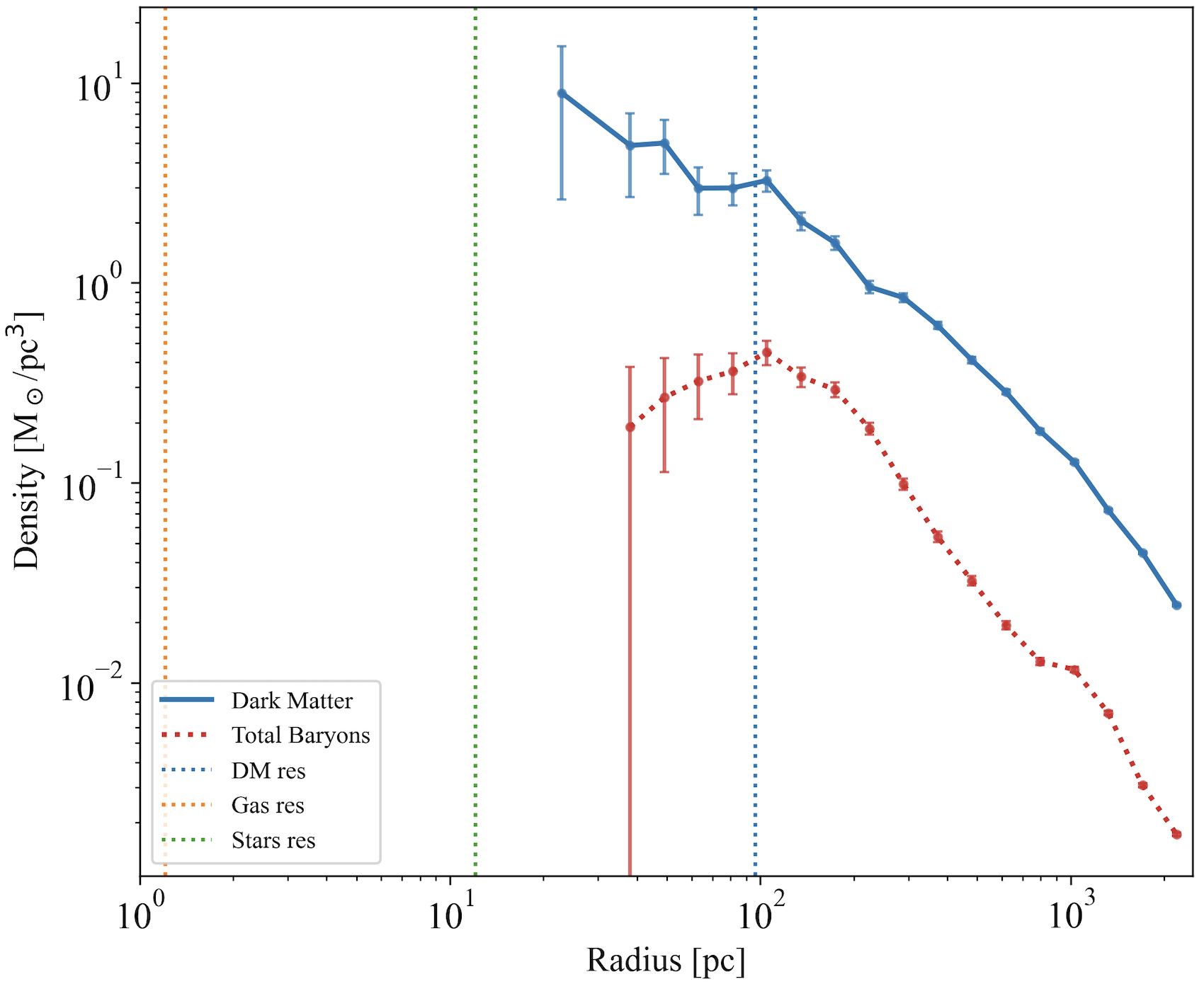}
    \includegraphics[width=.9\linewidth]{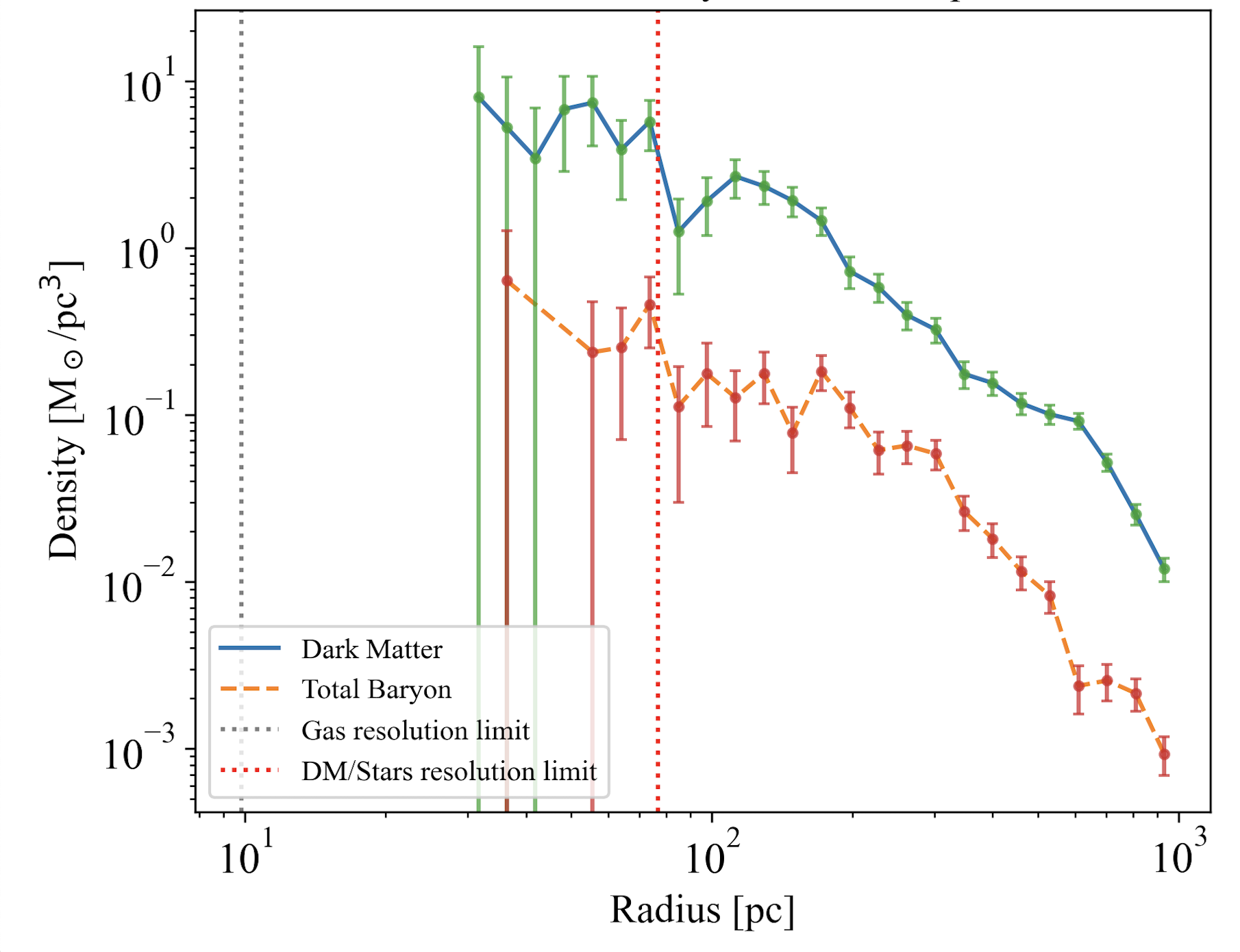}
    \caption{Radial density profiles of dark matter and total baryonic matter for the FIRE-2 A2 system at $z\approx10.6$ (upper panel) and the TNG50 system at $z\approx20$ (lower panel); error bars indicate Poisson uncertainties. Vertical dotted lines correspond to the spatial resolution scales set by the gravitational softening lengths of the different matter components.
   For the baryonic component, these correspond to minimum softening floors, since both the FIRE-2 and TNG50 simulations use adaptive baryonic resolution.
    }
    \label{den_prof}
\end{figure}

We have identified a number of systems in both the FIRE-2 and the TNG50 cosmological systems in which an SMBH appears at redshift $z=0$ and for which the initial massive seed black hole was formed at high redshift ($z \sim 7-10$).  As an illustration, the upper panel of Figure~\ref{den_prof} shows the radial density profiles of dark matter and total
baryonic matter at $z = 10.6$ for the primary halo of the FIRE-2 A2 zoom-in simulation  that  forms a central SMBH.   The bottom panel illustrates  density profiles for a similar system at  $z = 20$ that forms an SMBH within the TNG50 simulation. The point of this figure is to illustrate the density profiles just before the SMBH is seeded.

Note that the gas resolution limit shown in Figure~\ref{den_prof} at about 1 pc (upper figure) and 10 pc (lower panel) corresponds to the published minimum gas softening length converted to the corresponding redshift.  Both TNG50 and FIRE-2 use adaptive baryonic resolution, so the gas gravitational softening length varies locally rather than being globally fixed. The dark matter softening, on the other hand, is prescribed globally. 

However, a gas resolution limit of 1-10 pc does not mean the simulation is resolved  at a scale of 1-10 pc.   For example, the baryonic profile stops at the innermost non-empty radial shell. 

We also note that the innermost radii fall below the formal
\cite{Power2003} convergence limit in both simulations. The FIRE-2
comparison has a somewhat smaller formal convergence radius than the
corresponding TNG50 system, but in both cases the estimated convergence radius remains on the order of several hundred parsecs. 
 
We further note that in the FIRE-2 simulation, for example, the star formation criterion  requires a local hydrogen number density ($n_{\rm H} > 1000{\rm cm^{-3}}$) \citep{Hopkins_2018}, where $n_{\rm H}$ is the density assigned to individual gas resolution elements. In units of Figure 1, this would correspond to a gas mass density of about 
 32 M$_\odot$ pc$^{-3}$.  This is larger than the densities shown in Figure 1.   However, the profiles  shown in Figure 1  correspond to  spherically averaged baryonic density profiles computed over finite radial shells. The shell-averaged profile is of course expected to be smoother and lower than the local particle-scale density fluctuations in which the threshold used in the star formation prescription can be exceeded.

 Nevertheless, the resolved portions of the profiles can still be used to
explore plausible extrapolations of the DM distribution to smaller radii.
We consider both a Navarro-Frenk-White (NFW) profile \citep{nfw,klypin}
 of the form:
\begin{equation}
    \rho(r)=\frac{\rho_0 r_t}{r(1+r/r_t)^2},
\end{equation}
and a Pseudo-Jaffe (PJ) profile of the form as used in \cite{Powell25,Yu26} to fit the profile of collapsed DM perturbers:
\begin{equation}
    \rho(r)=\frac{\rho_0 r_t^4}{r^2(r^2+r_t^2)},
\end{equation}
 
 The best‐fitting parameters for the fit of the NFW profile to the TNG50 case are $(\rho_0,r_t) = (2.49 \pm 1.05~{\rm M}_\odot {\rm pc}^{-3}, 210.63 \pm 38.62~{\rm pc})$ for the dark matter density profile.  This implies a dark-matter density at small $r$ of $\rho(r) \approx (530 \pm 360) (1\text{ pc}/r)$  M$_\odot$ pc$^{-3}$.   For the Pseudo-Jaffe profile fit, the best‐fitting parameters are $(0.019 \pm 0.013~{\rm M}_\odot {\rm pc}^{-3}, 1163 \pm 332~{\rm pc})$, and the density at small $r$ is $\rho(r) \approx (2.6 \pm 1.4)\times 10^4\ (1\text{ pc}/r)^{2}$  M$_\odot$ pc$^{-3}$ for the dark matter density profile.  For the FIRE-2 simulation, the $r^{-2}$ profile corresponds to a density of $\rho(r) \sim (8.8 \pm 1.1)\times 10^4 \ (1\text{ pc}/r)^{2} ~{\rm M}_\odot {\rm pc}^{-3}$. 

Based upon both of these profiles, one could imagine an NSC based upon cooling gas within 100 pc on Figure \ref{den_prof} to a characteristic NSC radius of 0.2 pc.   This would imply a baryonic mass in the NSC of $\sim 10^5$ M$_\odot$ and a baryonic matter density of $\sim 10^8 {\rm ~M}_\odot {\rm ~pc}^{-3}$ characteristic of an NSC. If the dark matter retained an NFW profile, the total density of DM within the NSC would only be $\sim 10^5$ M$_\odot$ pc$^{-3}$, depending upon the lower integration limit.  However, DM clustered with the PJ profile would have a density  $\sim 10^8$ M$_\odot$ pc$^{-3}$.  Hence,  the DM density within the NSC could be comparable to or even greater than the baryonic mass density.
 Therefore, in this paper, we consider a range of possible DM content within an NSC that is comparable to the total baryonic density.  We then explore the impact of DM accretion on SMBH growth.

 \subsection{ULDM content}
We also examine the effect of ULDM on the growth of the SMBH. ULDM is often modeled as a free scalar field with a very small mass $\mu_s \sim 10^{-21}- 10^{-22}$ eV. Such a small mass implies a de Broglie wavelength consistent with the flattening of the DM density profile of some observed dwarf galaxies \cite{Hui17} on a scale of kpc.   In the non-relativistic limit, the configuration of such a field under the influence of gravity is governed by the Schrödinger–Poisson equations. The quasi-stationary, ground-state solutions of these equations are known as solitons, where quantum pressure exactly balances gravitational self-attraction. 

 The WKB approximation gives the solution to the Schrödinger–Poisson equations and yields a semi-analytic soliton density profile~\citep{Bucciotti_2023}: 
\small
\begin{equation}
4\pi \langle \rho \rangle \simeq
\begin{cases}
(8.3)\,\dfrac{M_{\mathrm{sol}}}{r_{\mathrm{sol}}^{3}}, & \text{for } r \to 0,\\[6pt]
(1.5)\,\dfrac{M_{\mathrm{sol}}}{r^{2} r_{\mathrm{sol}}}\,\sin^{2}\!\left(\dfrac{3\pi r}{4 r_{\mathrm{sol}}}\right), 
& \text{for } r \lesssim r_{\mathrm{sol}},\\[8pt]
(0.76)\,\dfrac{M_{\mathrm{sol}}}{r^{2} r_{\mathrm{sol}}}\,
\exp\!\left[-\,\dfrac{3\pi (r-r_{\mathrm{sol}})}{2 r_{\mathrm{sol}}}\right],
& \text{for } r \gtrsim r_{\mathrm{sol}},
\end{cases}
\label{eq:soliton-density}
\end{equation}
\normalsize
where \(M_{\mathrm{sol}}\) is the total soliton mass and \(r_{\mathrm{sol}}\) is the characteristic size of the soliton, which is associated with the mass of the scalar field  \(\mu_s\) through the following simplified argument:
According to the uncertainty principle, the quantum kinetic energy per particle is \(\sim1/(2\mu_sr_{\mathrm{sol}}^2)\), while the gravitational self potential energy per particle is \(\sim-GM_{\mathrm{sol}}u_s/r_{\mathrm{sol}}\). Balancing quantum pressure against self-gravity via the virial theorem yields \(GM_{\mathrm{sol}}r_{\mathrm{sol}}\mu_s^2\sim\mathcal{O}(1)\). A more careful solution of the Schrödinger–Poisson equations for the ground-state soliton gives a specific numerical constant \(\simeq6\). Therefore, \(r_{\mathrm{sol}}\simeq 6 (GM_{\mathrm{sol}}\mu_s^2)^{-1}\).

\begin{figure}[!h]
    \centering
    \includegraphics[width=1\linewidth]{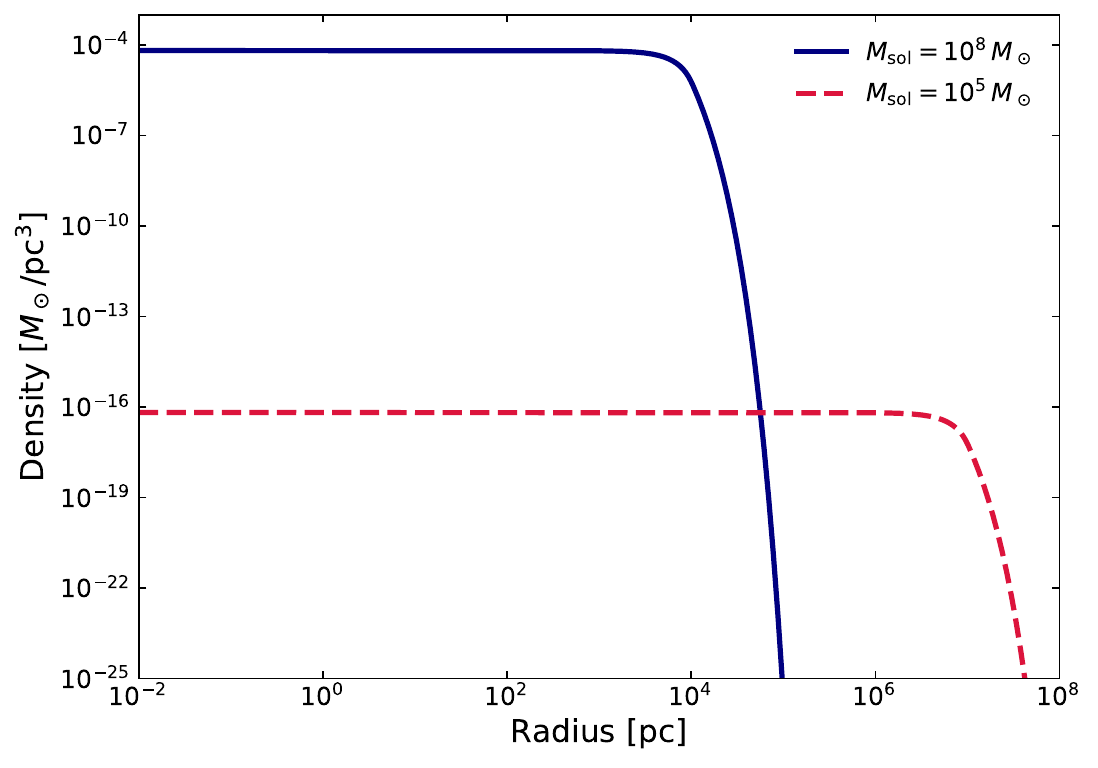}
    \caption{Soliton profile for a scalar field of mass $\mu_s = 10^{-22}$ eV associated with a model with total soliton mass \(M_{\mathrm{sol}}=10^5 \ \textup{M}_\odot\) and \(M_{\mathrm{sol}}=10^8 \ \textup{M}_\odot\).}
    \label{sol_prof}
\end{figure}

Figure \ref{sol_prof} shows an example of the soliton profile for an ULDM field of a typical mass of $10^{-22}$ eV and a total soliton mass of $10^8$ $\textup{M}_\odot$.  Under these conditions, we calculate the radius of the soliton core to be \(r_\text{sol}\simeq 10\) kpc and the total soliton mass to be $ \simeq  10^{8}$ M$_{\odot}$. Note that the soliton core radius is much larger than the half-mass radius of the NSC.  As described below, this causes differences in the SMBH growth when ULDM is included.  


In order to obtain an SMBH at the upper end of the $10^6 - 10^8$ M$_{\odot}$ range, we adopt models with a total NSC mass of $8 \times 10^8 $ M$_{\odot}$ \citep{MichaelSTurner_1991}. 
For baryonic matter, we took the initial star-to-gas mass ratio to be 4:1. We simulate clusters with initial half-mass radii of varying sizes (0.1, 0.2, 0.5, and 1.0 pc). These initial radii include values somewhat larger than those adopted by
\cite{Kritos2024} who only considered an initial half-mass radius of 0.1 pc.


\section{The Model} \label{sec:style}
The central BH grows through a combination of stellar BH mergers, dark-matter capture, and gas accretion, along with tidal disruption events with stars as described below and in \cite{Kritos2024}. 
The baryonic gas decreases exponentially over a characteristic gas expulsion timescale $\tau_{\text{ge}}$ and is taken to have a sound speed of 10 km s$^{-1}$.

The growth of the central black hole in the nuclear star cluster model considered here is described in part by the following set of coupled equations modified from those of \cite{Kritos2024}:
\begin{subequations} \label{1}
\begin{align}
\frac{d{\overline{m}_*}}{dt} &= -\nu \frac{\overline{m}_*}{t} \Theta(t - t_{\text{se}}) \label{1a}\\
\frac{dN_*}{dt} &= -\xi_e \frac{N_*}{\tau_{\text{rh}}} - R_{\text{TDE}} \label{1b}\\
\frac{dN_{\text{BH}}}{dt} &= -k_{\text{ejs}} R_{3\text{bb}} \label{1c}\\
\frac{dM_\text{g}}{dt} &= -\frac{M_\text{g}}{\tau_{\text{ge}}}- \dot{M}_{\text{acc}} \label{1d}\\
\frac{dM_{\text{BH}}}{dt} &= \dot{M}_{\text{acc}} + f_{\text{TDE}} \overline{m}_* R_{\text{TDE}} \label{1e}\\
&\quad +\rho_{\text{DM}} \langle\sigma_{\text{capt}} v_{\text{DM}}\rangle \notag\\
\frac{dr_{\text{h}}}{dt} &= \left( \frac{\zeta r_{\text{h}}}{\tau_{\text{rh}}} + \frac{2 r_{\text{h}}}{M_{\text{cl}}} \frac{dM_{\text{cl}}}{dt} \right) \Theta(t - \tau_{\text{cc}}) \label{1f}\\
&\quad - \frac{r_{\text{h}}}{M_{\text{cl}}} \left( N_* \frac{d\overline{m}_*}{dt} - \frac{M_\text{g}}{\tau_{\text{ge}}} \right) \notag
\end{align}
\end{subequations}

The first equation (\ref{1a}) describes the rate of change of the average stellar mass $\overline{m}_*$ after a stellar evolution time $t_{\text{se}} = 2$ Myr \citep{Alexander2014}, where $\Theta(t -t_{\text{se}})$ is the Heaviside step function. 
 The loss of massive stars at the end of their life will cause the averaged mass (integrated over the stellar IMF) to shift to smaller values in time. This evolution can be approximated by a power law \citep{Alexander2014}:
 \begin{equation}
     \overline{m}_* = \overline{m}_0 (t/\tau_{\text{se}})^{-\nu},
 \end{equation} 
 where $\nu = 0.07$.

The first term in the second equation (\ref{1b}) is an approximate fit to numerical simulations \citep{Alexander2014} of the loss of the number of stars $N_*$ due to relaxation-driven stellar escape. 
 Here, the half-mass relaxation time of the cluster $\tau_{\text{rh}}$ is the characteristic time over which a fraction $\psi$ of the total energy in the cluster is redistributed throughout the cluster via two-body relaxation. This can be approximated as 
 \begin{equation} \label{2}
     \tau_{\text{rh}} = 0.138 \sqrt{\frac{M_{\mathrm{cl}} r_\mathrm{h}^3}{G} } \frac{1}{{\overline{m}_*}\psi \ln{\Lambda}}~~,
 \end{equation}
with

 \begin{equation} \label{3}
     \psi = 1 + S_{\text{BH}}
 \end{equation} 
 where $S_{\text{BH}}$ is the Spitzer parameter given by

 \begin{equation} \label{4}
     S_{\text{BH}} = \left( \frac{N_{\text{BH}}}{N_{\text{star}}} \right) \times \left( \frac{m_{\text{BH}}}{m_{\text{star}}} \right)^{\frac{5}{2}}.
 \end{equation}

 The second term in Eq. \eqref{1b} is the loss of stars due to tidal disruption events. If one only considers the disruption of stars due to the central black hole, then the TDE rate is    
 \begin{equation} \label{5}
 R_{\text{TDE}} = \dot E_{\text{cl}}/Q_{\text{TDE}}~~,    
 \end{equation}
 where 
 \begin{equation} \label{6}
 Q_{\text{TDE}} = \frac{GM_{\text{BH}} {\overline{m}_*}}{2 r_\text{T}}    
 \end{equation} 
 is the heat released by each tidal disruption event and $r_\text{T}$ is the tidal radius.

The evolution of the number of stellar black holes in Eq. \eqref{1c} is governed by the ejection of BHs through three-body interactions. The parameter $k_{\rm ejs}$ is the efficiency of BH ejection by 3-body interactions, and $R_{3\text{bb}}$ is the rate of formation of the three-body systems \citep{antonini2020}.

The next equation (\ref{1d}) accounts for the loss of gas $M_\text{g}$ from the cluster through a characteristic gas expulsion timescale, $\tau_{\text{ge}}=100$ Myr, and accretion onto the central BH computed using an Eddington-limited Bondi accretion rate, $\dot{M}_{\text{acc}}$, as described below~\citep{Kritos2024}.

Eq. \eqref{1e} gives the adjusted mass accumulation rate of the central seed black hole due to gas accretion, stellar tidal disruption and dark matter capture.  The first term is the Eddington-limited Bondi rate, $\dot{M}_{\text{acc}}$, for gas accretion as discussed in \citep{Antonini2012} and \citep{Kritos2024}, i.e. for hydrogen gas accretion,  the maximum Bondi accretion rate is:
\begin{equation}
  \dot{M}_{\text{acc}} = \frac{4 \pi G M_{\rm BH} m_p} {\eta c \sigma_T} ~~,
\end{equation}
where $m_p$ is the proton mass, $\eta = 0.1$ is the radiative efficiency, and $\sigma_T$ is the Thomson scattering cross section. 
When gas accretion dominates the growth of the black hole mass, this leads to exponential growth $M_{\rm BH} = M_0 \exp\left( {t}/{\tau} \right)$ where
\begin{equation} 
\tau = \frac{\eta c \sigma_T}{4 \pi G m_p} \approx  50~{\rm Myr}~~,
\end{equation}
is the Salpeter timescale. 

The second term in Eq.~\eqref{1e} accounts for tidal disruption events.  Here, $f_{\text{TDE}}$ is the fraction of the stellar mass in stellar tidal disruption events that accrete onto the central BH. The variable $\overline{m}_{*}$ is the mean mass of stars, and $R_{\text{TDE}}$ is the rate of tidal disruption events. The third term describes the growth of the central BH by DM accretion in the cluster. The average is over the DM velocity distribution as described below.

 In addition to the above, the central BH growth rate is incremented at each time step from the BH mergers as described in \cite{Kritos2024}.   The number of 3-body BH systems is given by $R_{\rm 3bb} dt$.  These are sampled based upon a Poisson distribution and assigned masses and semi-major axes according to the initial hardness parameter of \cite{Morscher15}.  The remnant mass, spin and kick velocities are assigned according to the relativistic numerical simulations of \cite{Gerosa16}.  
 
 A fraction of these systems form a BH binary after merger.   As described in \cite{antonini2020} the simulation then checks whether the resulting binary has a larger gravitational radius $a_{\rm GW}$ than the ejection radius  $a_{\rm ej}$ \cite{antonini2020}, 
   where $a_{\rm GW}$ is the semi-major axis at which the probability in between binary and single interactions is unity.
   The ejection radius $a_{\rm ej}$ is the semi-major axis below which a three-body interaction will eject the binary from the cluster \citep{antonini2020}.
   
   If $a_{\rm GW} > a_{\rm ej}$, then the binary remains within the NSC.  If the kick velocity is less than the escape velocity, the mass of the central black hole M$_{\rm BH}$ is discretely increased by the mass of the binary remnant.  The black hole spin and number of BHs are also updated. Otherwise, the remnant is ejected.   

The last equation, (\ref{1f}), gives the rate of expansion of the half-mass radius \(r_{\mathrm{h}}\). It depends on two contributions. The first term is the expansion of the cluster from the half-mass relaxation timescale and mass change. It can be derived by directly differentiating the equation relating the total cluster energy \(E_{\mathrm{cl}}\), the total cluster mass \(M_{\mathrm{cl}}\), and the half-mass radius \(r_{\mathrm{h}}\) in virial equilibrium:
\begin{equation} \label{7}
    E_{\mathrm{cl}}=-\kappa\frac{GM_{\mathrm{cl}}^2}{2r_{\mathrm{h}}}~~,
\end{equation}
where $\kappa = 0.4$ for most spherical stellar systems~\citep{spitzer1969}, and using Hénon's principle~\citep{henon1961,henon1965, gieles2011}:
\begin{equation} \label{8}
    \frac{\dot{E}_{\mathrm{cl}}}{E_{\mathrm{cl}}}=-\frac{\zeta}{\tau_{\mathrm{rh}}},
\end{equation}
where $\tau_{\mathrm{rh}}$, defined by Eq.~\eqref{2}, is the characteristic timescale over which a fraction $\zeta \simeq 0.1$ of the total cluster energy can be transferred throughout the cluster via two-body relaxation. The quantity $\tau_{\mathrm{cc}}$ is the timescale for core collapse, {i.e.} the timescale for BHs to reach the center and start heating the cluster. This is taken as a fraction of the half-radius relaxation time, i.e. $\tau_{\mathrm{cc}} = 0.2 \tau_{\rm rh}$ \citep{Portegies02}.

 The second term in Eq. \eqref{1f} is due to adiabatic expansion after gas expulsion~\citep{hills1980, Kritos2024}.  It is derived from the successive disruption and restoration of virial equilibrium following each infinitesimal mass loss event experienced by the cluster, assuming the mass is ejected gradually over a time much longer than the dynamical time $\tau_{\mathrm{rh}}$. It is an adiabatic expansion term.  The expulsion of gas ejected by massive stars and residual gas within the cluster reduces its gravitational binding energy, thereby leading to cluster adiabatic expansion.

\subsection{Accretion of CDM}
We model cold dark matter (CDM) as a population of collisionless particles in a spherical potential,
with an absorbing sink at small angular momentum corresponding to direct capture by a centrally growing Schwarzschild
SMBH of mass $M_\mathrm{BH}$. 
Long-term capture is controlled by the loss-cone refilling in angular-momentum
space, rather than by a hydrodynamic (Bondi) inflow. 

The CDM capture rate is determined \cite{Shapiro83} by the minimum angular momentum, $J_{\rm min}(E) = {4 GM \over c}$, which defines a
loss-cone.
Specifically, the capture rate $\dot N$ of unbound collisionless particles is \citep{Shapiro83}:
\begin{equation}
    \dot N = 8 \pi^2 \int_0^\infty dE f(E) \int_0^{J_{\rm min}} dJ J 
\end{equation}
where $f(E)$ is the distribution function.
The growth rate of the black hole from particles of mass $m$ is then:
\begin{equation}
    \dot M = m \dot N  = \sigma_{\rm capt} \rho v_\infty
\end{equation}

 For a centrally growing Schwarzschild black hole of mass \(M_{\text{BH}}\) embedded in a stationary, isotropic bath of collisionless monoenergetic CDM particles with density \(\rho_{\text{DM}}\) and asymptotic speed $v_{\infty}\ll 1$ far from the black hole, the CDM accretion rate can be written as \citep{Shapiro83}:
\begin{equation} \label{9}
\sigma_{\text{capt}} = \frac{4\pi R_s^2}{v_{\infty}^2}, \qquad \dot M_{\text{BH}} = \rho_\text{DM}v_\infty\sigma_{\text{capt}} ,
\end{equation}
with \(R_s\equiv2GM_{\text{BH}}\). This applies to any non‑relativistic CDM candidate, whether WIMPs, primordial black holes (PBHs), sterile neutrinos, axions, etc.

In a realistic scenario, the CDM is collisionless with a
distribution function set by the complexities of the collapse process.  Nevertheless, since this is not known, it is reasonable to adopt a Maxwellian-Boltzmann velocity distribution \citep{Shapiro83}. For this case, the black-hole capture cross section increases by a numerical factor $ \sqrt{6 \pi}/\pi \sim 1.4$ \citep{Shapiro83}.  The fact that the capture rate is almost unchanged for either a monoenergetic or Maxwell-Boltzmann distribution suggests that there is not much sensitivity to the detailed distribution function.

 Another issue is that the distribution function can evolve once the accretion process begins.   In this work, we assume that the distribution function remains unchanged.  This is equivalent to assuming a very short relaxation time so that the distribution is immediately replenished as the CDM accretes onto the black hole.    In a future work, we will explore effects of the relaxation time on the DM capture rate.  However, since in the present work we are interested in the maximum rate of DM capture, we adopt a short relaxation time for the CDM distribution functions such that the loss cone has an extremely high replenishment efficiency.

As a side remark, we note that we are only concerned with the case of PBHs with a mass much smaller than that of the central accreting BH.  For low-mass PBHs (\(m_{\text{pbh}}\ll M_{\text{BH}}\)),  the formation of a bound binary through gravitational‑wave emission (GW capture) is strongly suppressed. The corresponding cross‑section is
\begin{equation} \label{10}
\sigma_{\text{GW}} = \frac{kG^2M^{12/7}\mu^{2/7}}{v_{\infty}^{18/7}c^{10/7}}\approx\frac{k R_s^2}{(v_{\infty}/c)^{18/7}}\left(\frac{m_{\text{pbh}}}{M_{\text{BH}}}\right)^{2/7},
\end{equation}
with \(M\equiv M_{\text{BH}} + m_{\text{pbh}}\), \(\mu\equiv {M_{\text{BH}} m_{\text{pbh}}}/(M_{\text{BH}} + m_{\text{pbh}})\), and \(k \equiv ({\pi}/{2}) (  {85 \pi}/{6\sqrt{2}} )^{2/7}\).  $\sigma_{\text{GW}}$ is significantly reduced by the small $(m_{\text{pbh}}/M_{\text{BH}})^{2/7}$ factor. Under any realistic halo velocity dispersion, this factor renders \(\sigma_{\text{GW}}\) several orders of magnitude smaller than \(\sigma_{\text{capt}}\).  Hence,  GW capture of low‑mass PBHs can be neglected in the DM accretion process considered here.

The CDM accretion rate will be dominated by the lowest velocities in the velocity distribution. To account for this, we assume that  unbound CDM obeys a Maxwell-Boltzmann velocity distribution far from the center BH, with mass density \(\rho_{\text{DM}}\) and root mean square (RMS) velocity given by the virial velocity:
\begin{equation} \label{11}
v_{\text{vir}} = \sqrt{\kappa\frac{GM_\mathrm{cl}}{r_\mathrm{h}}}~~,
\end{equation} 
{i.e.} the probability of finding DM of velocity $v_{\infty}$ within an interval $\mathrm{d}v_{\infty}$ is:
\begin{equation} \label{12}   
P(v_{\infty})\mathrm{d}v_{\infty}=4\pi^{-1/2}\frac{v_{\infty}^2}{\tilde{v}_{\text{vir}}^3}\exp{\left(-\frac{v_{\infty}^2}{\tilde{v}_\text{vir}^2}\right)}\mathrm{d}v_{\infty},
\end{equation}
where \(\tilde{v}_\text{vir}\equiv\sqrt{\frac{2}{3}}v_\text{vir}\).
Then, the CDM accretion rate can be calculated by
\begin{eqnarray} \label{13}
    \rho_{\text{DM}} \langle \sigma_{\text{capt}} v_{\infty}\rangle &=&\rho_{\text{DM}}\int_0^{\infty}\mathrm{d}v_{\infty} \sigma_{\text{capt}} v_{\infty}P(v_{\infty}) \nonumber  \\
    &=& 8 \sqrt{\pi} \rho_{\text{DM}} R_s^2 c^2 \tilde{v}_\text{vir}^{-1}~~.
\end{eqnarray}

The total mass of CDM in the NSC will decrease as particles are captured by the central black hole. The mass of CDM evolves in time according to:
\begin{equation} \label{14}
    \frac{dM_{\text{DM}}}{dt}=-\rho_{\text{DM}} \langle \sigma_{\text{capt}} v_{\infty}\rangle.
\end{equation}

\subsection{Accretion of ULDM}
In the case of ULDM, we take the evolution in BH mass due to the accretion of ULDM to be: 
\begin{equation}
    \frac{dM_{\text{BH,ULDM}}}{dt} =-\frac{dM_{\text{ULDM}}}{dt} =(0.04)R^2_{\text{s}}G^3M^4_{\text{sol}}\mu_s^6
\label{rateuldm}
\end{equation}
as derived in \cite{Bucciotti_2023}.  Here, $\mu_s$ is the mass of the scalar field, and we set $\hbar=c=1$. The density dependence is encoded in $M_{\text{sol}}$, which represents the total mass of ULDM that forms solitons enclosed within a radius \(r\) (based on the evolving soliton core radius) and is given by
\begin{equation}
    M_{\text{sol}} =  \int_0^r 4\pi r'^2 \, \rho_{\mathrm{sol}}\, \mathrm{d}r'.
\label{eq16}
\end{equation}

Following the derivation in \cite{Marsh}, the soliton mass and central density scale with the ULDM de Broglie wavelength as:
\begin{align}
M_{\mathrm{sol}} &\propto \lambda^{-1}, \\
\rho_{\mathrm{sol}} &\propto \lambda^{-4}.
\end{align}
Thus, 
\begin{equation}
\rho_{\mathrm{sol}} \propto M_{\mathrm{sol}}^4.
\end{equation}
This scaling is a consequence of the Schrödinger–Poisson equations that govern the ULDM soliton cores.
We adopt this relation to evolve the soliton density in time as:
\begin{equation}
\rho_{\mathrm{sol}}(t) = \rho_{\mathrm{sol},0} \left( \frac{M_{\mathrm{sol}}(t)}{M_{\mathrm{sol},0}} \right)^4.
\end{equation}




\section{Results}

To assess the role of DM in SMBH growth, we considered a variety of scenarios that incorporated DM accretion into the nuclear star cluster model.  We show that this modification significantly enhances the growth of a seed BH into an SMBH. 

\begin{figure}[]
    \centering
    \includegraphics[width=.8\linewidth]{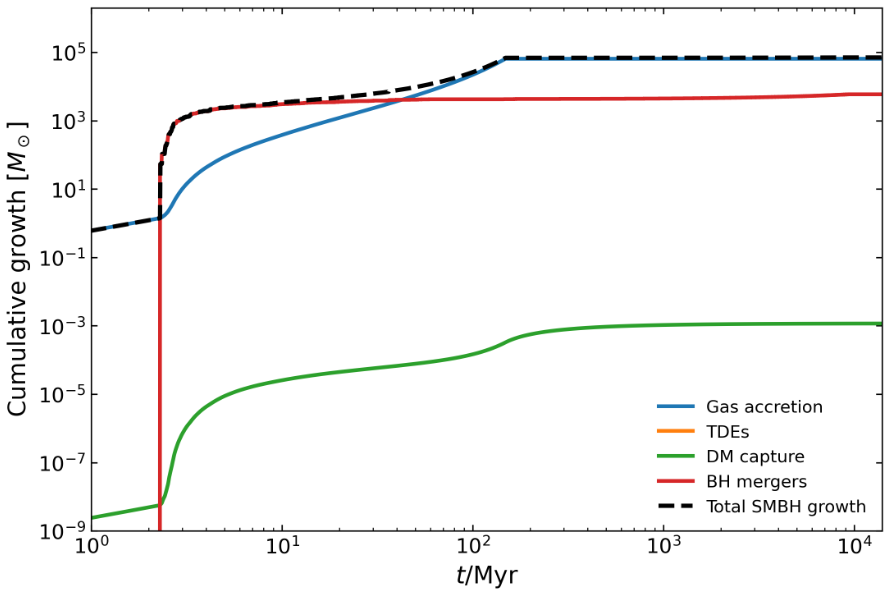}
    \includegraphics[width=.8\linewidth]{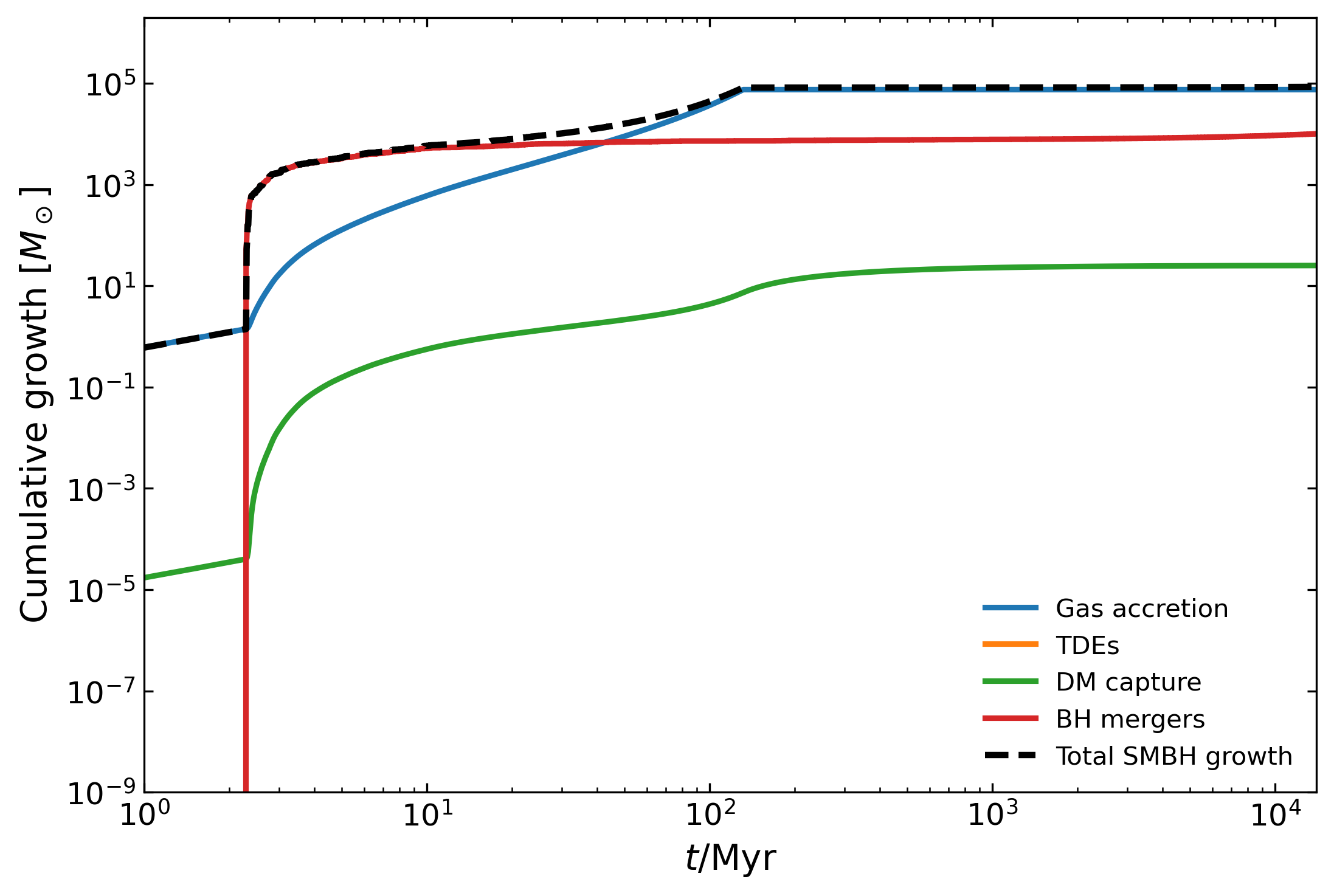}
    \caption{Contributions to the growth of an SMBH from a cluster of baryonic mass $10^{6}$ M$_\odot$ and DM content of 100 M$_\odot$ (top panel) and $10^{6}$ M$_\odot$ (bottom panel) as suggested by the NFW profile and Pseudo-Jaffe fit to data given in Fig. \ref{den_prof}. }
    \label{cdm_contributions}
\end{figure}

 As a first illustration, we consider a model motivated by the analysis of the TNG50 and FIRE simulations.  Fig.~\ref{cdm_contributions} shows the expected result if the CDM does not cluster, but instead maintains the NFW profile as baryonic matter cools to form the NSC with a radius of 0.2 pc.  The total cluster mass for this example was divided among DM, stars, black-hole remnants and gas. Stars and remnants were taken to constitute 80\%, while 20\% of the baryonic mass was in gas. 
 The initial mass in stars is then 80\% of the baryonic mass based upon the IMF.  This division of gas and stars is consistent with the models with a lower gas content studied in \cite{Kritos2024}.  We chose this division as a reasonable expectation after an initial burst of star formation. Also, lower gas content optimizes the possible influence of DM accretion which is what we wish to study in this paper. 
 
 In this case, the DM contributes only a minuscule amount to the final BH mass even when the DM mass is comparable to the initial mass in baryons.  For this relatively low-mass NSC, the BH growth is minimal (reaching only $10^5$ M$_\odot$) and what growth occurs is dominated by BH mergers and gas accretion. To model the formation of the observed M$_{\rm BH} > 10^7$ M$_\odot$ SMBH a more robust model must be considered.

Next, we therefore consider  more robust models similar to those studied in \cite{Kritos2024}.  We again begin with a seed black hole of 30 M$_{\odot}$.  However, in this case it is  embedded in an NSC of initial mass $8 \times 10^8$ M$_\odot$.  We consider uniform CDM and baryonic matter density profiles within an initial cluster with half-mass radii of 0.10, 0.20, 0.50, and 1 pc. Note that we consider models with larger initial half-mass radii than the 0.1 pc half-mass radius adopted in \cite{Kritos2024}.   

The total cluster mass for this example again was divided among DM, stars,  remnants and gas placing 20\% of the baryonic mass in gas.  However, we consider  a large DM-to-baryon density ratio of 5. 
 The initial mass in stars is then 80\% of the baryonic mass with stellar masses assigned according to the IMF.  

\begin{figure}[]
    \centering
    \includegraphics[width=.8\linewidth]{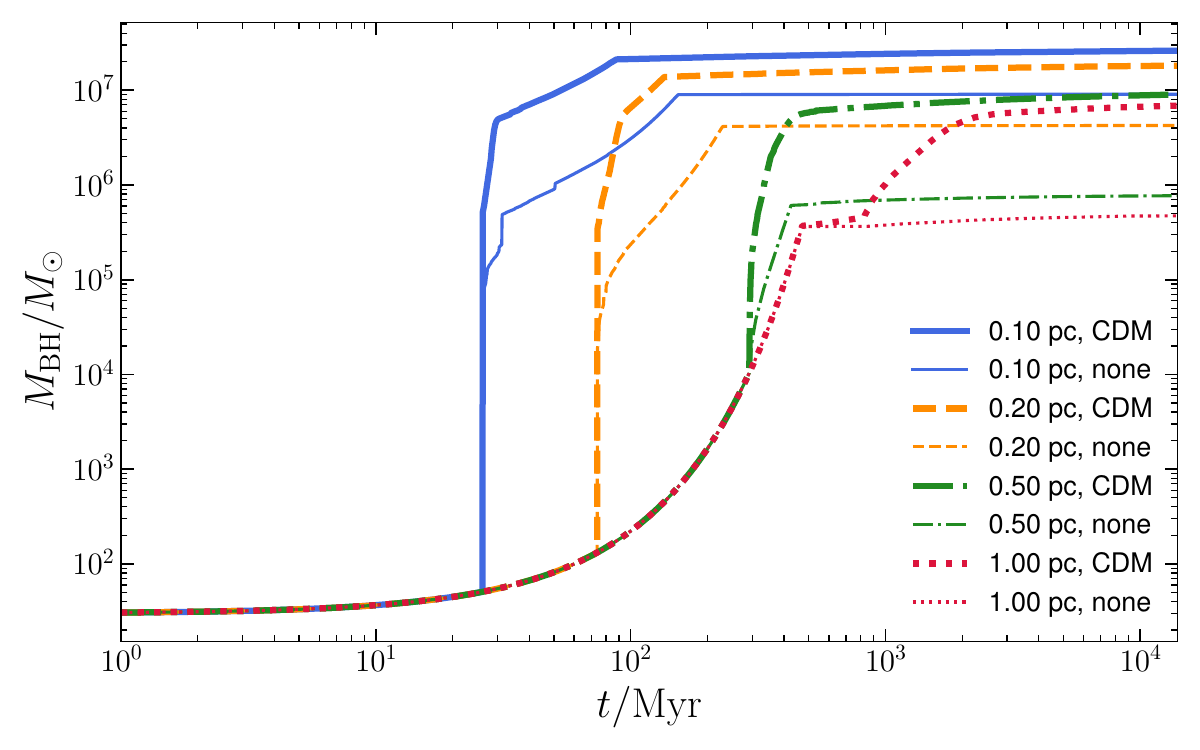}
    \includegraphics[width=.8\linewidth]{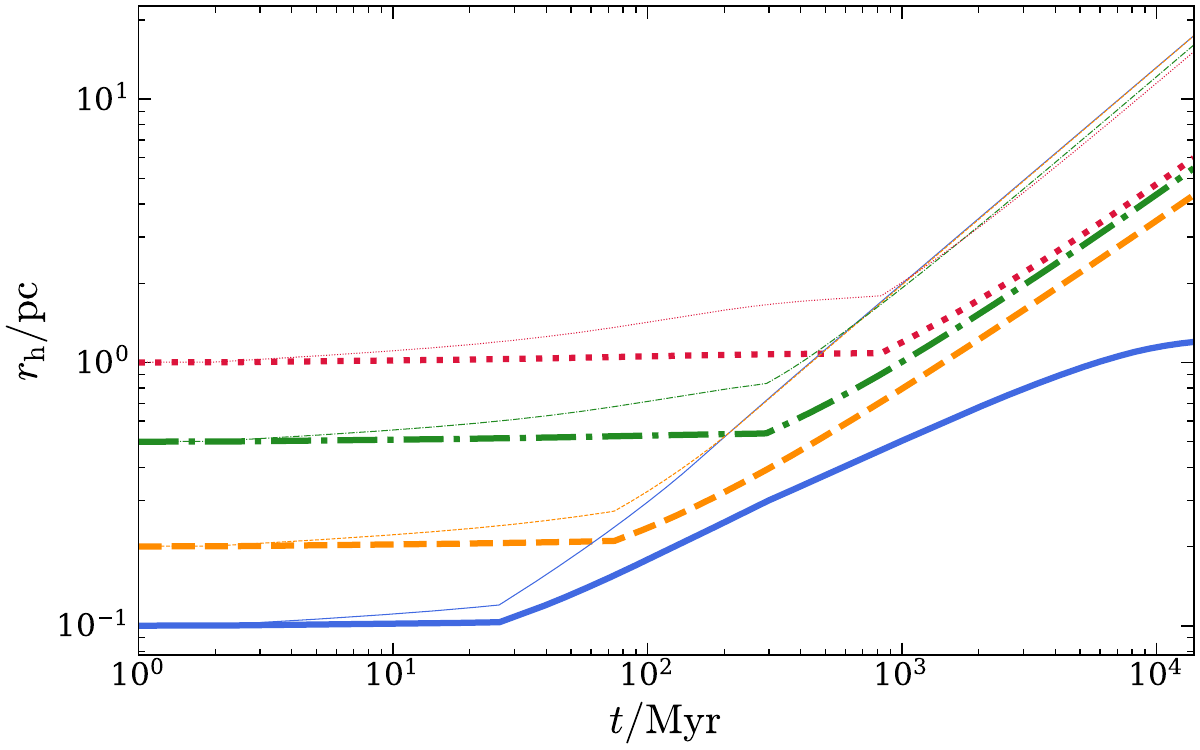}
    \includegraphics[width=.8\linewidth]{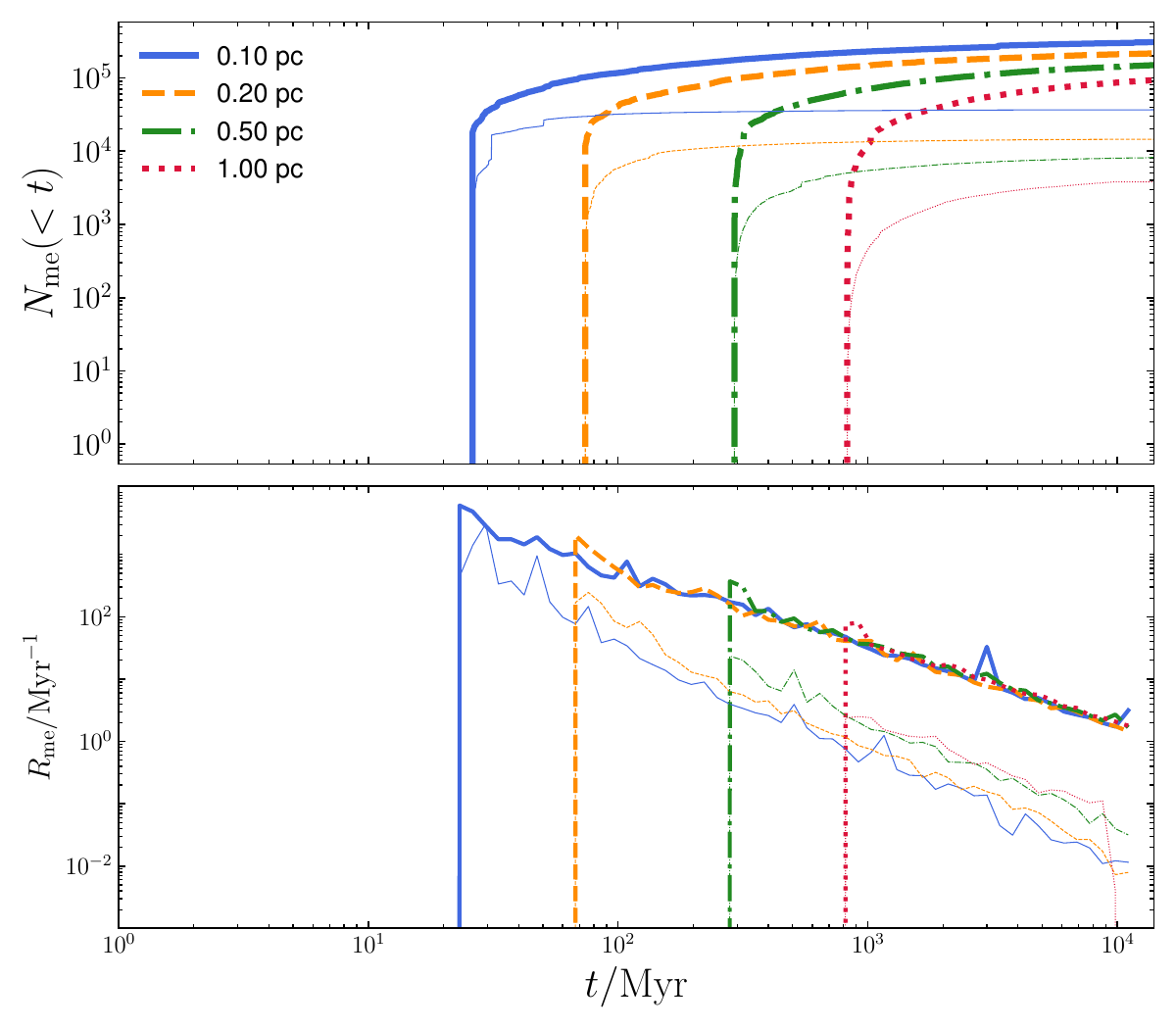}
    \includegraphics[width=.8\linewidth]{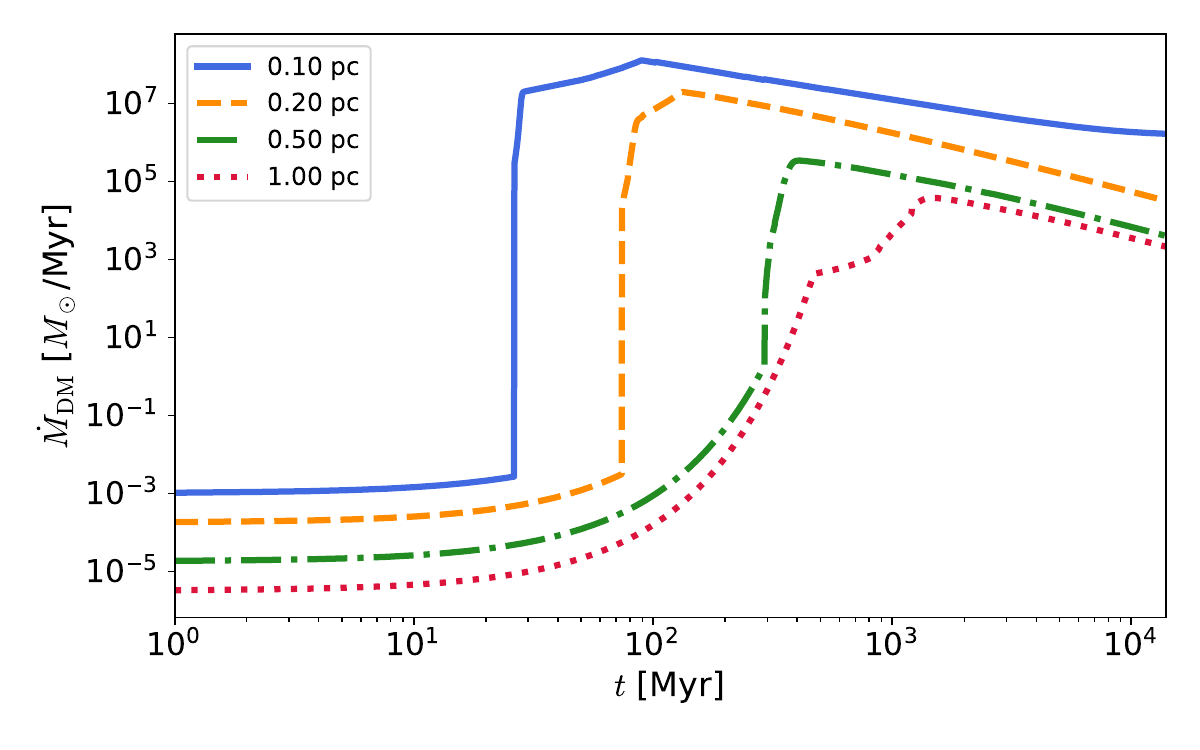}
    \caption{Top panel shows the evolution of the mass of the central black hole beginning with a 30 M$_{\odot}$ seed black hole within a cluster of mass $8 \times 10^8$ M$_\odot$. Lines correspond to  various initial half-mass radii as labeled. Lines are drawn with (CDM) and without (none) the presence of dark matter.  The second panel shows the growth of the half-mass radius of the NSC. The NSC begins to grow after the collapse time $\tau_{\rm cc}$ up to the observed sizes of $\sim$ $1 -10$ pc.  The third panel shows the burst of number of black hole mergers $N_{\rm me}$ once the collapse time has been exceeded.
    The fourth panel is the corresponding rate $R_{\rm me}$ of BH mergers. The bottom panel shows the rate of dark matter accretion $\dot M_{\rm DM}$.
    }
    \label{DM_panel}
\end{figure}

 The top panel of Figure \ref{DM_panel} shows the evolution of the mass of the SMBH for various initial half-mass radii and models with and without CDM.  All of these models exhibit an initial gradual growth of the central BH due to the combination of Eddington-limited Bondi accretion and a small amount of DM capture until the core collapse time $\tau_{\rm cc}$. 
 
 The growth of the half-mass radius $r_h$ is shown in the second panel of Figure \ref{DM_panel}.  The cluster half-mass radius  expands after $t =\tau_{\rm cc}$ according to Eq.~\eqref{1f}.    As noted in \cite{Kritos2024}, at $\tau_{\rm cc}$ the system is considered to be virialized and the seed black hole mass can rapidly grow by BH mergers as evidenced in the third and fourth panels of Fig.~\ref{DM_panel}.  Concurrently in our model, the rapid BH growth accelerates the capture of CDM after $\tau_{\rm cc}$ due to the dependence of the capture cross section on the BH mass.  This is illustrated by the evolution of $\dot M_{\rm DM}$ as shown in the bottom panel of Fig.~\ref{DM_panel}.
 
 From Eq.~\eqref{9}, the rate of mass of DM falling into the seed BH scales with the cross section, which in turn scales with the mass of the central black hole. 
\begin{equation} \label{15}
\sigma_{\text{capt}} \propto M_{\text{BH}}^{2}~~.
\end{equation}
Hence, once the seed BH becomes sufficiently large due to BH mergers, DM absorption proceeds rapidly. This leads to significant additional SMBH growth until the DM is depleted as shown in the bottom panel of Figure \ref{DM_panel}.

Figure \ref{DM_panel} shows that the most dense cluster with an initial virial half-mass radius of 0.10 parsecs attains an SMBH of $\sim 3 \times 10^7$ M$_{\odot}$ within 200 Myr, nearly accounting for the observation \citep{Natarajan24} of an SMBH of $4 \times 10^7$ M$_\odot$ at $z = 10$ and easily  explaining the presence of an SMBH of mass $1.5 \times 10^{7}$ M$_\odot$ in a near-pristine galaxy at z = 7.04 as described by \cite{Maiolino2025}.  In contrast, the simulation without CDM only produces an SMBH of $\leq 10^7$ M$_\odot$. Hence, the presence of CDM enhances the early growth of the mass of the  SMBH by a factor of $\sim 3$, thereby providing the dominant contribution to the final mass of the SMBH.

The lower density cluster with a 0.2 pc initial half-mass radius reaches $\sim 2\times 10^7$ M$_{\odot}$  within approximately 100 Myr. In this case, the simulations without DM only lead to a BH mass of $\sim 4\times 10^6$ M$_\odot$.  Hence, in this case  the presence of DM enhances the mass of the SMBH by a factor of $\sim 5$. The NSC  with an $r_h =$ 0.5 pc  reaches $\sim 10^7$ M$_{\odot}$  within approximately 1000 Myr. In this case, the simulations without DM do not produce an SMBH $> 8 \times10^5$ M$_{\odot}$ within $10^4$ Myr. Hence, the DM in this case increases the SMBH mass by an order of magnitude.

The case with an initial half-mass radius of 1 pc exhibits a somewhat different evolution.  The accretion of gas and DM are depleted after about 400 Myr.  However, $\tau_{\rm cc}$ is not achieved until $t\sim 1000$ Myr.  Hence, for this model the onset of BH mergers and additional DM capture appears as a second growth beginning at 1000 Myr after a plateau in BH mass forms at 400 Myr.


\begin{figure}
    \centering
    \includegraphics[width=1\linewidth]{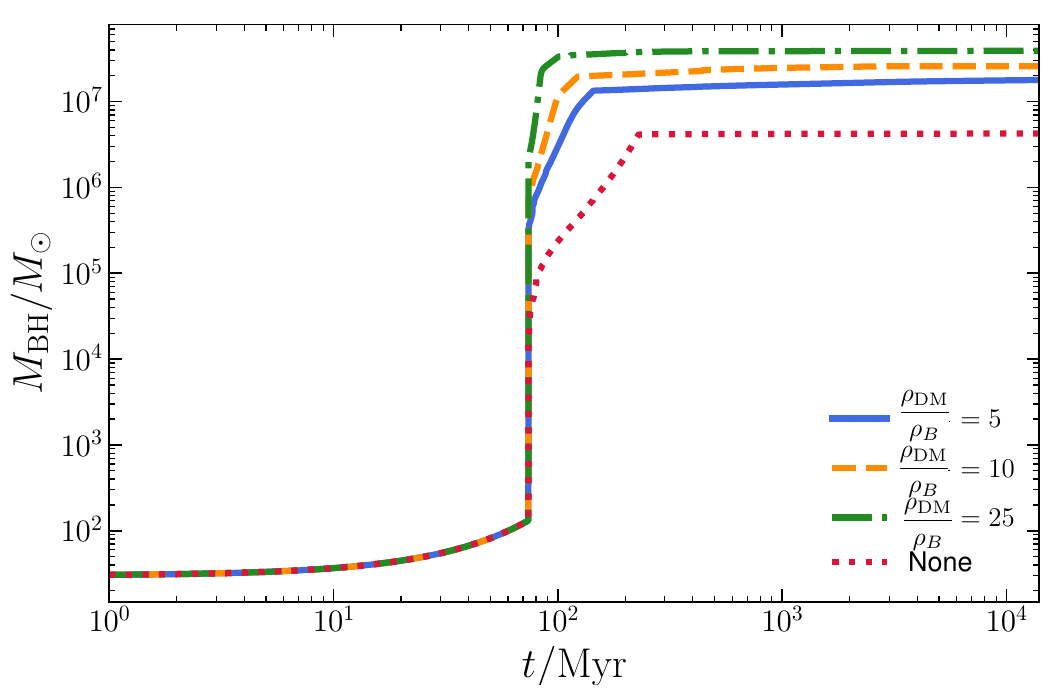}
    \caption{SMBH growth in models with varying DM-to-baryon ratios while maintaining a constant baryon cluster mass.  Other parameters are the same as in Figure \ref{DM_panel} for the model with an initial half-mass radius of 0.20 pc.}
    \label{dm_vary}
\end{figure}

 In Figure \ref{dm_vary} we consider the possibility of even larger CDM clustering and examine the dependence of SMBH growth on the  DM-to-baryon ratio. 
This figure shows that higher  DM/Baryonic ratios accelerate the formation of the SMBH and lead to a higher SMBH mass. The final SMBH mass is roughly proportional to the DM/Baryon ratio, i.e. increasing the ratio from 5 to 10 increases the SMBH mass from $1.8\times 10^7$ to $2.6\times 10^7$, while increasing the ratio by a factor of 5 increases the SMBH mass from $1.8\times 10^7$ to $4 \times 10^7$.

\subsection{Models with ULDM}
\begin{figure}
    \centering
    \includegraphics[width=1\linewidth]{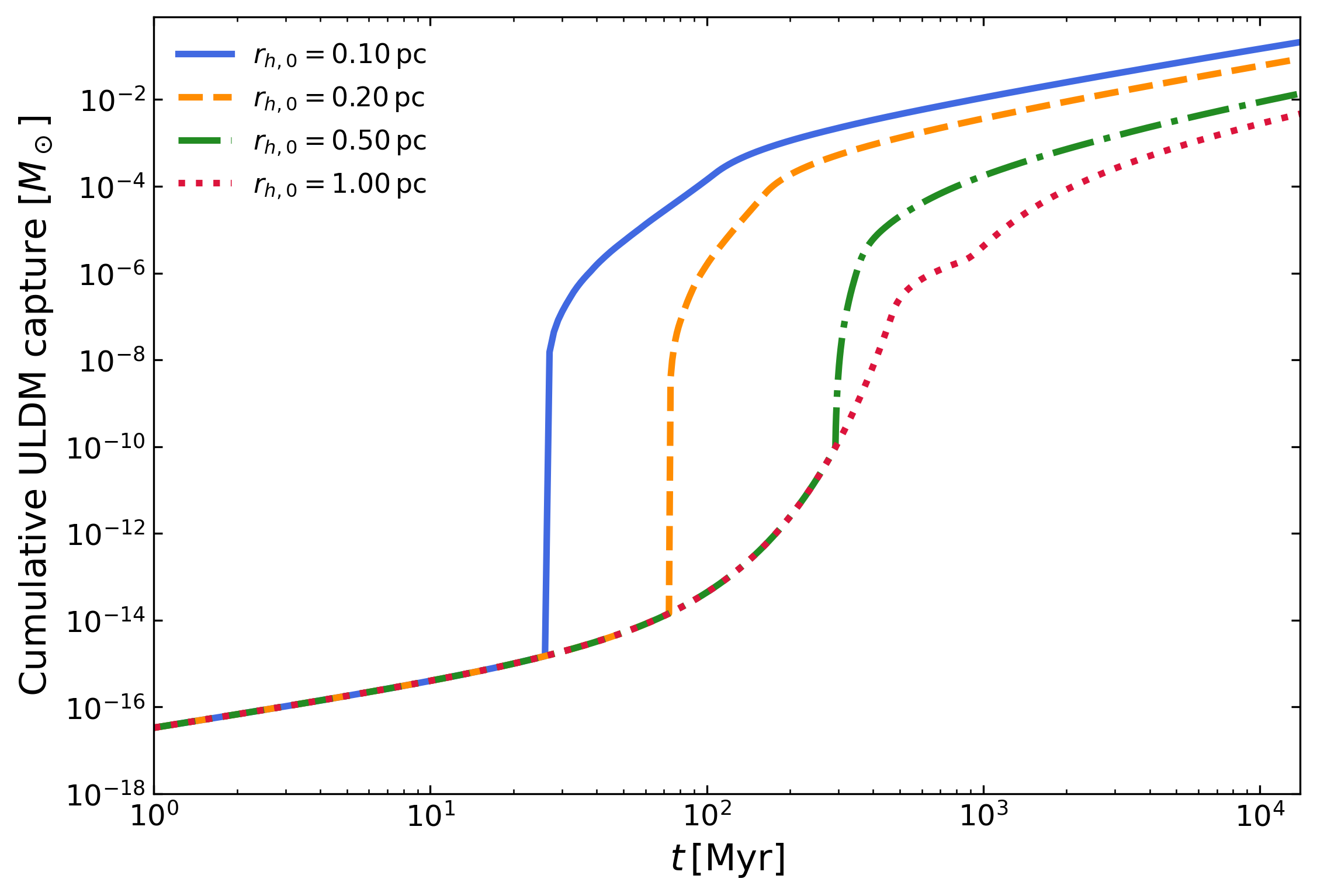}
    \caption{ULDM capture for scalar field mass of $\mu_s = 10^{-22}$ eV with varying initial half-mass radii. The initial conditions are the same as in Figure \ref{DM_panel}.}
    \label{uldm_contrubtion}
\end{figure}

\begin{figure}
    \centering
    \includegraphics[width=1\linewidth]{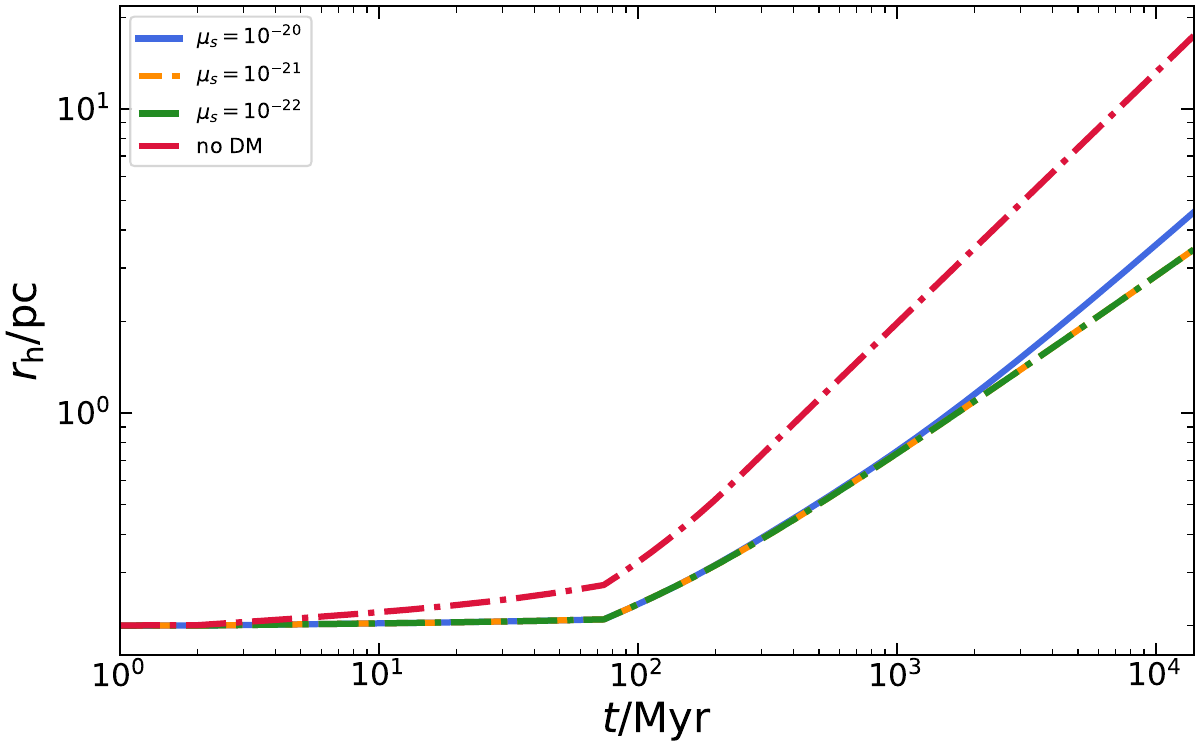}
     \caption{Growth of the half-mass radius of the NSC with various values for the ULDM mass and $r_h = 0.2$ pc. }
    \label{uldm_rh}
\end{figure}

\begin{figure}
    \centering
    \includegraphics[width=1\linewidth]{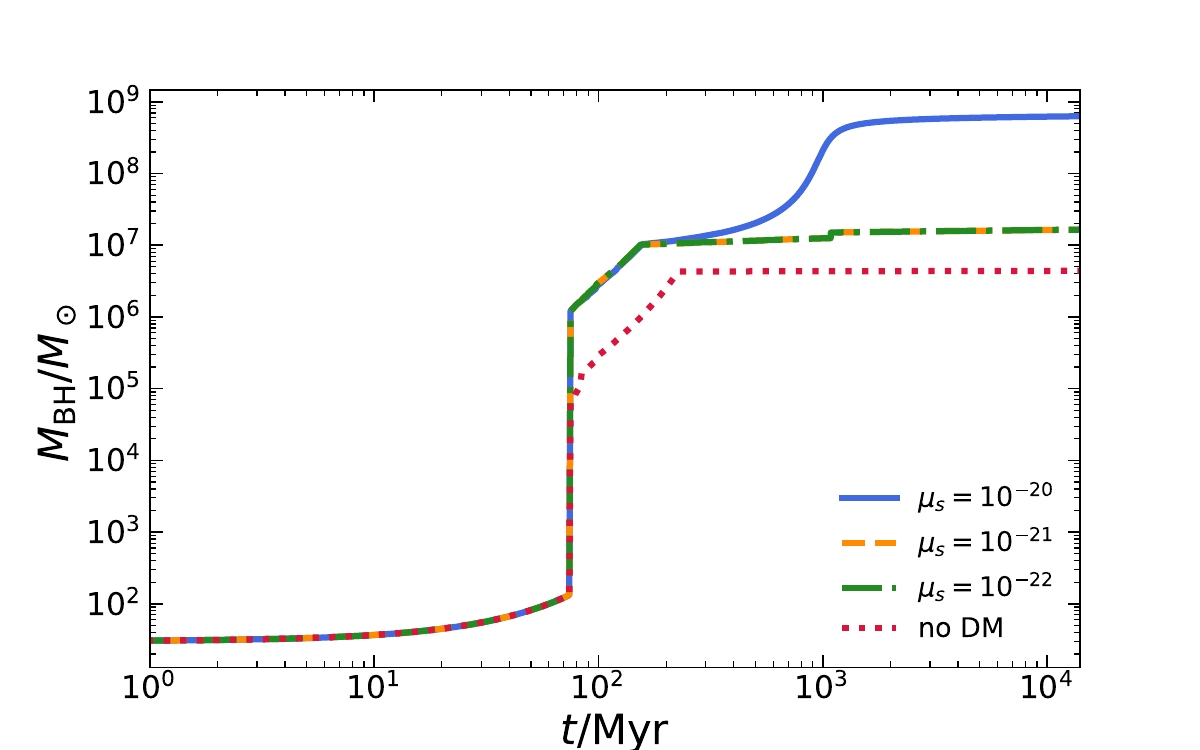}
    \caption{Models with various masses $\mu_s$ for the scalar field ultralight DM. The initial conditions are the same as in Figure \ref{DM_panel} with an initial half-mass radius of 0.2 pc.}
    \label{dif_scal}
\end{figure}

\begin{figure}
    \centering
    \includegraphics[width=1\linewidth]{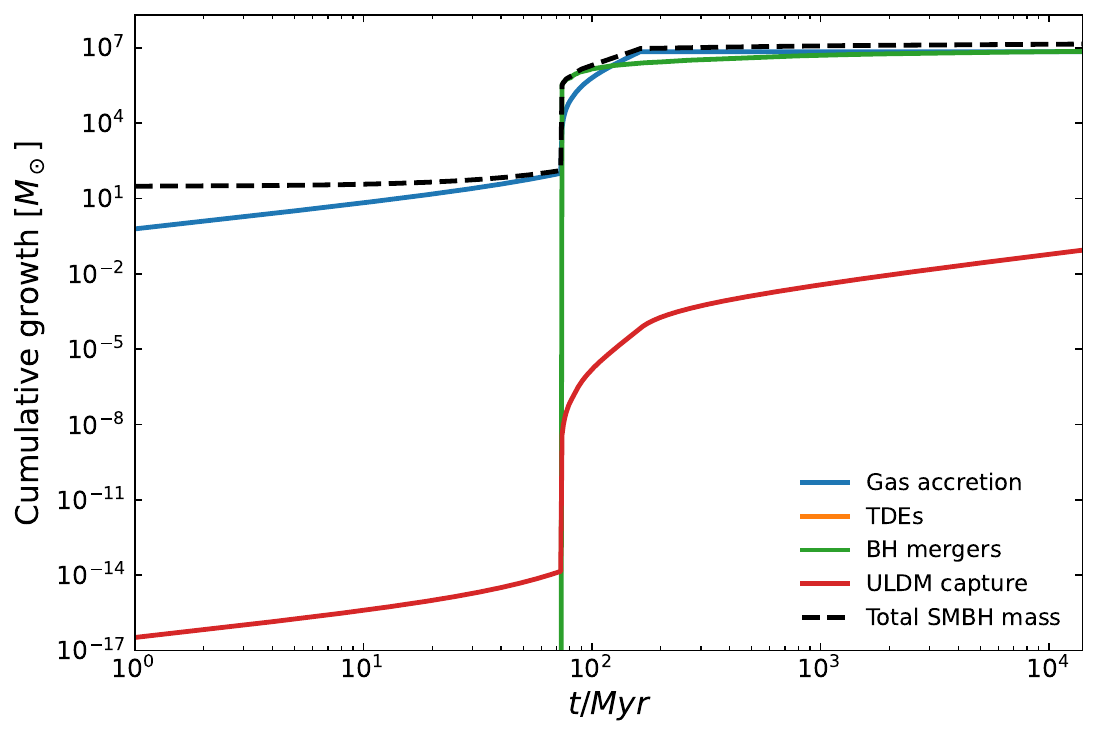}
    \caption{Contributions to the final SMBH mass in the case of ULDM with a scalar field mass of $\mu_s$ = $10^{-22}$ eV and half-mass radius of r$_h$ = 0.20 pc.}
    \label{uldm_22_contribution}
\end{figure}

Figures \ref{uldm_contrubtion}, \ref{uldm_rh}, and \ref{uldm_22_contribution} illustrate the evolution of SMBH growth in the case of ULDM. Figure \ref{dif_scal} shows the growth of the central black hole for the case of the most massive scalar field considered here, $\mu_s = 10^{-22}$ eV. This figure shows that ULDM can have an even more dramatic increase in the SMBH mass.  For $\mu_s = 10^{-22}$ eV, the growth of the central BH occurs in two stages. First, there is a rapid SMBH growth rate due to BH mergers once $t > \tau_{\rm cc}$. This is shown in Figures \ref{dif_scal} and \ref{cdm_contributions}.   This occurs for all models with or without ULDM.  

For the $\mu_s = 10^{-22}$ eV model, however, there is also a second burst of SMBH growth occurring at $t \ge 1$ Gyr.  We attribute this to the following:
The gas and stellar BH accretion essentially stops at 200-300 Myr. This can be seen in the no-DM curve and also the cases of lighter ULDM mass as shown in Figure \ref{dif_scal}.  This is due to the expansion of the cluster and diminishing rate of BH mergers.

However, the accretion of ULDM continues because the density of the soliton is constant even though the core radius is expanding.   
The soliton density remains constant due to its large de Broglie wavelength as shown in Figure \ref{sol_prof} and the middle panel of Figure \ref{DM_panel}. 
Hence, the ULDM accretion remains robust until the core radius exceeds the soliton radius $\sim 1$ pc after about 1 Gyr, after which $\rho_{sol}$ decreases rapidly, leading to a decrease in $dM_{\text{BH}}/dt$, and the SMBH stops growing as shown in the top and bottom panels of Figure \ref{DM_panel}.

This highlights a difference in the physics of BH accretion for ULDM.  Since the ULDM de Broglie wavelength of the ultralight particle can be comparable to the size of the core radius, e.g. Fig.~\ref{sol_prof}, the density remains constant as the core radius grows.  

The SMBH growth rate due to ULDM, however,  depends sensitively on the scalar field mass, i.e. $\propto \mu_s^6$ as in Eq.~(\ref{rateuldm}).  Figure \ref{dif_scal} shows the growth of the central black hole for various masses of the ultralight scalar particle and an initial core radius of 0.2 pc. For the case of smaller scalar field masses, the second burst of BH growth is suppressed by the $\mu^6$ factor in Eq.~(\ref{eq16}). This is true even though the soliton core is much larger for the lighter ULDM masses.
Hence, for lighter ULDM masses only the initial burst of BH mergers contributes to the growth of the SMBH.

\section{Conclusions}
 In summary, we have studied a  parameter set of nuclear star cluster initial conditions that include possible effects of dark matter accretion onto an SMBH.   We show that in a scenario in which the DM distribution is unaffected by the cooling and collapse of gas to an NSC, the effect of DM is minimal on the growth of an SMBH even for a cuspy NFW profile.  However, we have also considered  models in which the DM could be clustered with $\rho(r) \propto r^{-2}$ on small scales as inferred for some disrupters.    In this context, we have considered a broad range of possible DM content within nuclear star clusters.  

We find that with the inclusion of CDM (or ULDM with mass $\mu_s \sim 10^{-22}$ eV), with a density  greater than that of baryonic matter in the NSC, a central seed black hole can much more easily grow to the $4 \times 10^7$ M$_\odot$  SMBH detected at $z=10$ by JWST \citep{Natarajan24} and easily grows to $1.5 \times 10^{7}$ M$_\odot$ SMBH by z = 7.04 as described by \cite{Maiolino2025}.

Indeed, depending upon the ratio of DM to baryonic matter, the dark matter can enhance the mass of the SMBH by an order of magnitude or more. We also highlight here the possible unique contribution of ULDM capture at late times due to its large de Broglie wavelength that keeps the soliton density large until the NSC radius expands beyond the size of the de Broglie wavelength of the ULDM.




\begin{acknowledgements}
    Work at the University of Notre Dame is supported by the U.S. Department of Energy under Nuclear Theory Grant DE-FG02-95-ER40934. We acknowledge support from the University of Notre Dame Center for Research Computing. A. I. thanks Lauren Weiss for helpful discussions on simulation development related to this work.
\end{acknowledgements}


\bibliography{references}{}

@article{Yu26,
  title = {Core-Collapsed SIDM Halos as the Common Origin of Dense Perturbers in Lenses, Streams, and Satellites},
  author = {Yu, Hai-Bo},
  journal = {Phys. Rev. Lett.},
  volume = {136},
  issue = {14},
  pages = {141001},
  numpages = {5},
  year = {2026},
  month = {Apr},
  publisher = {American Physical Society},
  doi = {10.1103/txxx-97ln},
  url = {https://link.aps.org/doi/10.1103/txxx-97ln}
}

@ARTICLE{Powell25,
       author = {{Powell}, D.~M. and {McKean}, J.~P. and {Vegetti}, S. and {Spingola}, C. and {White}, S.~D.~M. and {Fassnacht}, C.~D.},
        title = "{A million-solar-mass object detected at a cosmological distance using gravitational imaging}",
      journal = {Nature Astronomy},
         year = 2025,
        month = nov,
       volume = {9},
        pages = {1714-1722},
          doi = {10.1038/s41550-025-02651-2},
archivePrefix = {arXiv},
       eprint = {2510.07382},
 primaryClass = {astro-ph.CO},
       adsurl = {https://ui.adsabs.harvard.edu/abs/2025NatAs...9.1714P}
}

@ARTICLE{Bonaca19,
       author = {{Bonaca}, Ana and {Hogg}, David W. and {Price-Whelan}, Adrian M. and {Conroy}, Charlie},
        title = "{The Spur and the Gap in GD-1: Dynamical Evidence for a Dark Substructure in the Milky Way Halo}",
      journal = {\apj},
         year = 2019,
        month = jul,
       volume = {880},
       number = {1},
          eid = {38},
        pages = {38},
          doi = {10.3847/1538-4357/ab2873},
archivePrefix = {arXiv},
       eprint = {1811.03631},
 primaryClass = {astro-ph.GA},
       adsurl = {https://ui.adsabs.harvard.edu/abs/2019ApJ...880...38B}
}

@ARTICLE{Nibauer25,
       author = {{Nibauer}, Jacob and {Bonaca}, Ana and {Price-Whelan}, Adrian M. and {Spergel}, David N. and {Greene}, Jenny E.},
        title = "{Measurement of Dark Matter Substructure from the Kinematics of the GD-1 Stellar Stream}",
      journal = {arXiv e-prints},
         year = 2025,
        month = oct,
          eid = {arXiv:2510.02247},
        pages = {arXiv:2510.02247},
          doi = {10.48550/arXiv.2510.02247},
archivePrefix = {arXiv},
       eprint = {2510.02247},
 primaryClass = {astro-ph.GA},
       adsurl = {https://ui.adsabs.harvard.edu/abs/2025arXiv251002247N}
}

@ARTICLE{Hui17,
       author = {{Hui}, Lam and {Ostriker}, Jeremiah P. and {Tremaine}, Scott and {Witten}, Edward},
        title = "{Ultralight scalars as cosmological dark matter}",
      journal = {\prd},
         year = 2017,
        month = feb,
       volume = {95},
       number = {4},
          eid = {043541},
        pages = {043541},
          doi = {10.1103/PhysRevD.95.043541},
archivePrefix = {arXiv},
       eprint = {1610.08297},
 primaryClass = {astro-ph.CO},
       adsurl = {https://ui.adsabs.harvard.edu/abs/2017PhRvD..95d3541H}
}

@BOOK{Shapiro83,
       author = {{Shapiro}, Stuart L. and {Teukolsky}, Saul A.},
        title = "{Black holes, white dwarfs and neutron stars. The physics of compact objects}",
         year = 1983,
           publisher={John Wiley \& Sons},
             isbn={0-471-87317-9},
          doi = {10.1002/9783527617661},
       adsurl = {https://ui.adsabs.harvard.edu/abs/1983bhwd.book.....S}
}

@ARTICLE{Natarajan24,
       author = {{Natarajan}, Priyamvada and {Pacucci}, Fabio and {Ricarte}, Angelo and {Bogd{\'a}n}, {\'A}kos and {Goulding}, Andy D. and {Cappelluti}, Nico},
        title = "{First Detection of an Overmassive Black Hole Galaxy UHZ1: Evidence for Heavy Black Hole Seed Formation from Direct Collapse}",
      journal = {\apjl},
         year = 2024,
        month = jan,
       volume = {960},
       number = {1},
          eid = {L1},
        pages = {L1},
          doi = {10.3847/2041-8213/ad0e76},
archivePrefix = {arXiv},
       eprint = {2308.02654},
 primaryClass = {astro-ph.HE},
       adsurl = {https://ui.adsabs.harvard.edu/abs/2024ApJ...960L...1N}
}

@article{Nelson:2019,
   title={First results from the TNG50 simulation: galactic outflows driven by supernovae and black hole feedback},
   volume={490},
   ISSN={1365-2966},
   url={http://dx.doi.org/10.1093/mnras/stz2306},
   DOI={10.1093/mnras/stz2306},
   number={3},
   journal={Monthly Notices of the Royal Astronomical Society},
   publisher={Oxford University Press (OUP)},
   author={Nelson, Dylan and Pillepich, Annalisa and Springel, Volker and Pakmor, Rüdiger and Weinberger, Rainer and Genel, Shy and Torrey, Paul and Vogelsberger, Mark and Marinacci, Federico and Hernquist, Lars},
   year={2019},
   month=aug, pages={3234–3261} }

@article{Pillepich:2019,
	adsurl = {https://ui.adsabs.harvard.edu/abs/2019MNRAS.490.3196P},
	archiveprefix = {arXiv},
	author = {{Pillepich}, Annalisa and {Nelson}, Dylan and {Springel}, Volker and {Pakmor}, R{\"u}diger and {Torrey}, Paul and {Weinberger}, Rainer and {Vogelsberger}, Mark and {Marinacci}, Federico and {Genel}, Shy and {van der Wel}, Arjen and {Hernquist}, Lars},
	doi = {10.1093/mnras/stz2338},
	eprint = {1902.05553},
	journal = {\mnras},
	month = dec,
	number = {3},
	pages = {3196-3233},
	primaryclass = {astro-ph.GA},
	title = {{First results from the TNG50 simulation: the evolution of stellar and gaseous discs across cosmic time}},
	volume = {490},
	year = 2019}

@ARTICLE{Morscher15,
       author = {{Morscher}, Meagan and {Pattabiraman}, Bharath and {Rodriguez}, Carl and {Rasio}, Frederic A. and {Umbreit}, Stefan},
        title = "{The Dynamical Evolution of Stellar Black Holes in Globular Clusters}",
      journal = {\apj},
         year = 2015,
        month = feb,
       volume = {800},
       number = {1},
          eid = {9},
        pages = {9},
          doi = {10.1088/0004-637X/800/1/9},
archivePrefix = {arXiv},
       eprint = {1409.0866},
 primaryClass = {astro-ph.GA},
       adsurl = {https://ui.adsabs.harvard.edu/abs/2015ApJ...800....9M}
}

@ARTICLE{Planck-Collaboration:2016,
       author = {{Planck Collaboration} and {Ade}, P.~A.~R. and {Aghanim}, N. and {Arnaud}, M. and {Ashdown}, M. and {Aumont}, J. and {Baccigalupi}, C. and {Banday}, A.~J. and {Barreiro}, R.~B. and {Bartlett}, J.~G. and {Bartolo}, N. and {Battaner}, E. and {Battye}, R. and {Benabed}, K. and {Beno{\^\i}t}, A. and {Benoit-L{\'e}vy}, A. and {Bernard}, J. -P. and {Bersanelli}, M. and {Bielewicz}, P. and {Bock}, J.~J. and {Bonaldi}, A. and {Bonavera}, L. and {Bond}, J.~R. and {Borrill}, J. and {Bouchet}, F.~R. and {Boulanger}, F. and {Bucher}, M. and {Burigana}, C. and {Butler}, R.~C. and {Calabrese}, E. and {Cardoso}, J. -F. and {Catalano}, A. and {Challinor}, A. and {Chamballu}, A. and {Chary}, R. -R. and {Chiang}, H.~C. and {Chluba}, J. and {Christensen}, P.~R. and {Church}, S. and {Clements}, D.~L. and {Colombi}, S. and {Colombo}, L.~P.~L. and {Combet}, C. and {Coulais}, A. and {Crill}, B.~P. and {Curto}, A. and {Cuttaia}, F. and {Danese}, L. and {Davies}, R.~D. and {Davis}, R.~J. and {de Bernardis}, P. and {de Rosa}, A. and {de Zotti}, G. and {Delabrouille}, J. and {D{\'e}sert}, F. -X. and {Di Valentino}, E. and {Dickinson}, C. and {Diego}, J.~M. and {Dolag}, K. and {Dole}, H. and {Donzelli}, S. and {Dor{\'e}}, O. and {Douspis}, M. and {Ducout}, A. and {Dunkley}, J. and {Dupac}, X. and {Efstathiou}, G. and {Elsner}, F. and {En{\ss}lin}, T.~A. and {Eriksen}, H.~K. and {Farhang}, M. and {Fergusson}, J. and {Finelli}, F. and {Forni}, O. and {Frailis}, M. and {Fraisse}, A.~A. and {Franceschi}, E. and {Frejsel}, A. and {Galeotta}, S. and {Galli}, S. and {Ganga}, K. and {Gauthier}, C. and {Gerbino}, M. and {Ghosh}, T. and {Giard}, M. and {Giraud-H{\'e}raud}, Y. and {Giusarma}, E. and {Gjerl{\o}w}, E. and {Gonz{\'a}lez-Nuevo}, J. and {G{\'o}rski}, K.~M. and {Gratton}, S. and {Gregorio}, A. and {Gruppuso}, A. and {Gudmundsson}, J.~E. and {Hamann}, J. and {Hansen}, F.~K. and {Hanson}, D. and {Harrison}, D.~L. and {Helou}, G. and {Henrot-Versill{\'e}}, S. and {Hern{\'a}ndez-Monteagudo}, C. and {Herranz}, D. and {Hildebrandt}, S.~R. and {Hivon}, E. and {Hobson}, M. and {Holmes}, W.~A. and {Hornstrup}, A. and {Hovest}, W. and {Huang}, Z. and {Huffenberger}, K.~M. and {Hurier}, G. and {Jaffe}, A.~H. and {Jaffe}, T.~R. and {Jones}, W.~C. and {Juvela}, M. and {Keih{\"a}nen}, E. and {Keskitalo}, R. and {Kisner}, T.~S. and {Kneissl}, R. and {Knoche}, J. and {Knox}, L. and {Kunz}, M. and {Kurki-Suonio}, H. and {Lagache}, G. and {L{\"a}hteenm{\"a}ki}, A. and {Lamarre}, J. -M. and {Lasenby}, A. and {Lattanzi}, M. and {Lawrence}, C.~R. and {Leahy}, J.~P. and {Leonardi}, R. and {Lesgourgues}, J. and {Levrier}, F. and {Lewis}, A. and {Liguori}, M. and {Lilje}, P.~B. and {Linden-V{\o}rnle}, M. and {L{\'o}pez-Caniego}, M. and {Lubin}, P.~M. and {Mac{\'\i}as-P{\'e}rez}, J.~F. and {Maggio}, G. and {Maino}, D. and {Mandolesi}, N. and {Mangilli}, A. and {Marchini}, A. and {Maris}, M. and {Martin}, P.~G. and {Martinelli}, M. and {Mart{\'\i}nez-Gonz{\'a}lez}, E. and {Masi}, S. and {Matarrese}, S. and {McGehee}, P. and {Meinhold}, P.~R. and {Melchiorri}, A. and {Melin}, J. -B. and {Mendes}, L. and {Mennella}, A. and {Migliaccio}, M. and {Millea}, M. and {Mitra}, S. and {Miville-Desch{\^e}nes}, M. -A. and {Moneti}, A. and {Montier}, L. and {Morgante}, G. and {Mortlock}, D. and {Moss}, A. and {Munshi}, D. and {Murphy}, J.~A. and {Naselsky}, P. and {Nati}, F. and {Natoli}, P. and {Netterfield}, C.~B. and {N{\o}rgaard-Nielsen}, H.~U. and {Noviello}, F. and {Novikov}, D. and {Novikov}, I. and {Oxborrow}, C.~A. and {Paci}, F. and {Pagano}, L. and {Pajot}, F. and {Paladini}, R. and {Paoletti}, D. and {Partridge}, B. and {Pasian}, F. and {Patanchon}, G. and {Pearson}, T.~J. and {Perdereau}, O. and {Perotto}, L. and {Perrotta}, F. and {Pettorino}, V. and {Piacentini}, F. and {Piat}, M. and {Pierpaoli}, E. and {Pietrobon}, D. and {Plaszczynski}, S. and {Pointecouteau}, E. and {Polenta}, G. and {Popa}, L. and {Pratt}, G.~W. and {Pr{\'e}zeau}, G.},
        title = "{Planck 2015 results. XIII. Cosmological parameters}",
      journal = {\aap},
         year = 2016,
        month = sep,
       volume = {594},
          eid = {A13},
        pages = {A13},
          doi = {10.1051/0004-6361/201525830},
archivePrefix = {arXiv},
       eprint = {1502.01589},
 primaryClass = {astro-ph.CO},
       adsurl = {https://ui.adsabs.harvard.edu/abs/2016A&A...594A..13P}
}

@ARTICLE{Portegies02,
       author = {{Portegies Zwart}, Simon F. and {McMillan}, Stephen L.~W.},
        title = "{The Runaway Growth of Intermediate-Mass Black Holes in Dense Star Clusters}",
      journal = {\apj},
         year = 2002,
        month = sep,
       volume = {576},
       number = {2},
        pages = {899-907},
          doi = {10.1086/341798},
archivePrefix = {arXiv},
       eprint = {astro-ph/0201055},
 primaryClass = {astro-ph},
       adsurl = {https://ui.adsabs.harvard.edu/abs/2002ApJ...576..899P}
}

@ARTICLE{Gerosa16,
       author = {{Gerosa}, Davide and {Kesden}, Michael},
        title = "{precession: Dynamics of spinning black-hole binaries with python}",
      journal = {\prd},
         year = 2016,
        month = jun,
       volume = {93},
       number = {12},
          eid = {124066},
        pages = {124066},
          doi = {10.1103/PhysRevD.93.124066},
archivePrefix = {arXiv},
       eprint = {1605.01067},
 primaryClass = {astro-ph.HE},
       adsurl = {https://ui.adsabs.harvard.edu/abs/2016PhRvD..93l4066G}
}

@ARTICLE{Neumayer2020,
       author = {{Neumayer}, Nadine and {Seth}, Anil and {B{\"o}ker}, Torsten},
        title = "{Nuclear star clusters}",
      journal = {\aapr},
         year = 2020,
        month = jul,
       volume = {28},
       number = {1},
          eid = {4},
        pages = {4},
          doi = {10.1007/s00159-020-00125-0},
archivePrefix = {arXiv},
       eprint = {2001.03626},
 primaryClass = {astro-ph.GA},
       adsurl = {https://ui.adsabs.harvard.edu/abs/2020A&ARv..28....4N}
}

@article{Devecchi2010,
    author = "Devecchi, B. and Volonteri, M. and Colpi, M. and Haardt, F.",
    title = "{High redshift formation and evolution of central massive objects I: model description}",
    eprint = "1001.3874",
    archivePrefix = "arXiv",
    primaryClass = "astro-ph.CO",
    doi = "10.1111/j.1365-2966.2010.17363.x",
    journal = "Mon. Not. Roy. Astron. Soc.",
    volume = "409",
    pages = "1057",
    year = "2010"
}

@article{Vanzella2023,
   title={JWST/NIRCam Probes Young Star Clusters in the Reionization Era
					Sunrise Arc},
   volume={945},
   ISSN={1538-4357},
   url={http://dx.doi.org/10.3847/1538-4357/acb59a},
   DOI={10.3847/1538-4357/acb59a},
   number={1},
   journal={The Astrophysical Journal},
   publisher={American Astronomical Society},
   author={Vanzella, Eros and Claeyssens, Adélaïde and Welch, Brian and Adamo, Angela and Coe, Dan and Diego, Jose M. and Mahler, Guillaume and Khullar, Gourav and Kokorev, Vasily and Oguri, Masamune and Ravindranath, Swara and Furtak, Lukas J. and Hsiao, Tiger Yu-Yang and Abdurro’uf and Mandelker, Nir and Brammer, Gabriel and Bradley, Larry D. and Bradač, Maruša and Conselice, Christopher J. and Dayal, Pratika and Nonino, Mario and Andrade-Santos, Felipe and Windhorst, Rogier A. and Pirzkal, Nor and Sharon, Keren and de Mink, S. E. and Fujimoto, Seiji and Zitrin, Adi and Eldridge, Jan J. and Norman, Colin},
   year={2023},
   month=mar, pages={53} }

@ARTICLE{Kritos2024,
       author = {{Kritos}, Konstantinos and {Berti}, Emanuele and {Silk}, Joseph},
        title = "{Supermassive black holes from runaway mergers and accretion in nuclear star clusters}",
      journal = {\mnras},
         year = 2024,
        month = jun,
       volume = {531},
       number = {1},
        pages = {133-136},
          doi = {10.1093/mnras/stae1145},
archivePrefix = {arXiv},
       eprint = {2404.11676},
 primaryClass = {astro-ph.HE},
       adsurl = {https://ui.adsabs.harvard.edu/abs/2024MNRAS.531..133K}
}

@ARTICLE{Begelman2006,
       author = {{Begelman}, Mitchell C. and {Volonteri}, Marta and {Rees}, Martin J.},
        title = "{Formation of supermassive black holes by direct collapse in pre-galactic haloes}",
      journal = {\mnras},
         year = 2006,
        month = jul,
       volume = {370},
       number = {1},
        pages = {289-298},
          doi = {10.1111/j.1365-2966.2006.10467.x},
archivePrefix = {arXiv},
       eprint = {astro-ph/0602363},
 primaryClass = {astro-ph},
       adsurl = {https://ui.adsabs.harvard.edu/abs/2006MNRAS.370..289B}
}

@article{Latif2022,
   title={Turbulent cold flows gave birth to the first quasars},
   volume={607},
   ISSN={1476-4687},
   url={http://dx.doi.org/10.1038/s41586-022-04813-y},
   DOI={10.1038/s41586-022-04813-y},
   number={7917},
   journal={Nature},
   publisher={Springer Science and Business Media LLC},
   author={Latif, M. A. and Whalen, D. J. and Khochfar, S. and Herrington, N. P. and Woods, T. E.},
   year={2022},
   month=jul, pages={48–51} }

@article{Volonteri2008,
  author = {Volonteri, Marta and Lodato, Giuseppe and Natarajan, Priyamvada},
  title = {The Evolution of Massive Black Hole Seeds},
  journal = {Monthly Notices of the Royal Astronomical Society},
  volume = {383},
  number = {3},
  pages = {1079-1088},
  year = {2008},
  doi = {10.1111/j.1365-2966.2007.12589.x},
  publisher = {Oxford University Press},
  url = {https://doi.org/10.1111/j.1365-2966.2007.12589.x}
}

@article{Xu_2013,
doi = {10.1088/0004-637X/773/2/83},
url = {https://dx.doi.org/10.1088/0004-637X/773/2/83},
year = {2013},
month = {jul},
publisher = {The American Astronomical Society},
volume = {773},
number = {2},
pages = {83},
author = {Hao Xu and John H. Wise and Michael L. Norman},
title = {POPULATION III STARS AND REMNANTS IN HIGH-REDSHIFT GALAXIES},
journal = {The Astrophysical Journal},
}

@article{MichaelSTurner_1991,
doi = {10.1088/0031-8949/1991/T36/018},
url = {https://dx.doi.org/10.1088/0031-8949/1991/T36/018},
year = {1991},
month = {jan},
publisher = {},
volume = {1991},
number = {T36},
pages = {167},
author = {Michael S Turner},
title = {Dark matter in the Universe},
journal = {Physica Scripta},
}

@article{Alexander2014,
   title={A prescription and fast code for the long-term evolution of star clusters – III. Unequal masses and stellar evolution},
   volume={442},
   ISSN={0035-8711},
   url={http://dx.doi.org/10.1093/mnras/stu899},
   DOI={10.1093/mnras/stu899},
   number={2},
   journal={Monthly Notices of the Royal Astronomical Society},
   publisher={Oxford University Press (OUP)},
   author={Alexander, Poul E. R. and Gieles, Mark and Lamers, Henny J. G. L. M. and Baumgardt, Holger},
   year={2014},
   month=jun, pages={1265–1285} }

@ARTICLE{Antonini2012,
       author = {{Antonini}, Fabio},
        title = "{Origin and Growth of Nuclear Star Clusters around Massive Black Holes}",
      journal = {\apj},
         year = 2013,
        month = jan,
       volume = {763},
       number = {1},
          eid = {62},
        pages = {62},
          doi = {10.1088/0004-637X/763/1/62},
archivePrefix = {arXiv},
       eprint = {1207.6589},
 primaryClass = {astro-ph.GA},
       adsurl = {https://ui.adsabs.harvard.edu/abs/2013ApJ...763...62A}
}

@article{Antonini2020,
   title={Population synthesis of black hole binary mergers from star clusters},
   volume={492},
   ISSN={1365-2966},
   url={http://dx.doi.org/10.1093/mnras/stz3584},
   DOI={10.1093/mnras/stz3584},
   number={2},
   journal={MNRAS},
   publisher={Oxford University Press (OUP)},
   author={Antonini, Fabio and Gieles, Mark},
   year={2020},
   month=jan, pages={2936–2954} }

@ARTICLE{nfw,
       author = {{Navarro}, Julio F.},
        title = "{The Inner Structure of Cold Dark Matter Halos}",
      journal = {arXiv e-prints},
         year = 2001,
        month = oct,
          eid = {astro-ph/0110680},
        pages = {astro-ph/0110680},
          doi = {10.48550/arXiv.astro-ph/0110680},
archivePrefix = {arXiv},
       eprint = {astro-ph/0110680},
 primaryClass = {astro-ph},
       adsurl = {https://ui.adsabs.harvard.edu/abs/2001astro.ph.10680N}
}

@ARTICLE{klypin,
       author = {{Klypin}, Anatoly and {Zhao}, HongSheng and {Somerville}, Rachel S.},
        title = "{{\ensuremath{\Lambda}}CDM-based Models for the Milky Way and M31. I. Dynamical Models}",
      journal = {\apj},
         year = 2002,
        month = jul,
       volume = {573},
       number = {2},
        pages = {597-613},
          doi = {10.1086/340656},
archivePrefix = {arXiv},
       eprint = {astro-ph/0110390},
 primaryClass = {astro-ph},
       adsurl = {https://ui.adsabs.harvard.edu/abs/2002ApJ...573..597K}
}

@ARTICLE{gieles2011,
       author = {{Gieles}, Mark and {Heggie}, Douglas C. and {Zhao}, Hongsheng},
        title = "{The life cycle of star clusters in a tidal field}",
      journal = {\mnras},
         year = 2011,
        month = jun,
       volume = {413},
       number = {4},
        pages = {2509-2524},
          doi = {10.1111/j.1365-2966.2011.18320.x},
archivePrefix = {arXiv},
       eprint = {1101.1821},
 primaryClass = {astro-ph.GA},
       adsurl = {https://ui.adsabs.harvard.edu/abs/2011MNRAS.413.2509G}
}

@article{henon1961,
  title={Sur l'{\'e}volution dynamique des amas globulaires},
  author={H{\'e}non, Michel},
  journal={Annales d'Astrophysique, Vol. 24, p. 369},
  volume={24},
  pages={369},
  year={1961}
}

@article{henon1965,
  title={Sur l'{\'e}volution dynamique des amas globulaires. II. Amas isol{\'e}},
  author={H{\'e}non, M},
  journal={Annales d'Astrophysique, Vol. 28, p. 62},
  volume={28},
  pages={62},
  year={1965}
}

@ARTICLE{hills1980,
       author = {{Hills}, J.~G.},
        title = "{The effect of mass loss on the dynamical evolution of a stellar system - Analytic approximations}",
      journal = {\apj},
         year = 1980,
        month = feb,
       volume = {235},
        pages = {986-991},
          doi = {10.1086/157703},
       adsurl = {https://ui.adsabs.harvard.edu/abs/1980ApJ...235..986H}
}

@article{Biernacki_2017,
   title={On the dynamics of supermassive black holes in gas-rich, star-forming galaxies: the case for nuclear star cluster co-evolution},
   volume={469},
   ISSN={1365-2966},
   url={http://dx.doi.org/10.1093/mnras/stx845},
   DOI={10.1093/mnras/stx845},
   number={1},
   journal={Monthly Notices of the Royal Astronomical Society},
   publisher={Oxford University Press (OUP)},
   author={Biernacki, Pawel and Teyssier, Romain and Bleuler, Andreas},
   year={2017},
   month=apr, pages={295–313} }

@article{Susa_2014,
   title={THE MASS SPECTRUM OF THE FIRST STARS},
   volume={792},
   ISSN={1538-4357},
   url={http://dx.doi.org/10.1088/0004-637X/792/1/32},
   DOI={10.1088/0004-637x/792/1/32},
   number={1},
   journal={The Astrophysical Journal},
   publisher={American Astronomical Society},
   author={Susa, Hajime and Hasegawa, Kenji and Tominaga, Nozomu},
   year={2014},
   month=aug, pages={32} }

@article{Davies_2011,
   title={SUPERMASSIVE BLACK HOLE FORMATION VIA GAS ACCRETION IN NUCLEAR STELLAR CLUSTERS},
   volume={740},
   ISSN={2041-8213},
   url={http://dx.doi.org/10.1088/2041-8205/740/2/L42},
   DOI={10.1088/2041-8205/740/2/l42},
   number={2},
   journal={The Astrophysical Journal},
   publisher={American Astronomical Society},
   author={Davies, Melvyn B. and Coleman Miller, M. and Bellovary, Jillian M.},
   year={2011},
   month=sep, pages={L42} }

@ARTICLE{liempi2024,
       author = {{Liempi}, M. and {Schleicher}, D.~R.~G. and {Benson}, A. and {Escala}, A. and {Vergara}, M.~C.},
        title = "{The supermassive black hole population from seeding via collisions in nuclear star clusters}",
      journal = {\aap},
         year = 2025,
        month = feb,
       volume = {694},
          eid = {A42},
        pages = {A42},
          doi = {10.1051/0004-6361/202451672},
archivePrefix = {arXiv},
       eprint = {2412.08280},
 primaryClass = {astro-ph.GA},
       adsurl = {https://ui.adsabs.harvard.edu/abs/2025A&A...694A..42L}
}

@ARTICLE{Kroupa02,
       author = {{Kroupa}, Pavel},
        title = "{The Initial Mass Function of Stars: Evidence for Uniformity in Variable Systems}",
      journal = {Science},
         year = 2002,
        month = jan,
       volume = {295},
       number = {5552},
        pages = {82-91},
          doi = {10.1126/science.1067524},
archivePrefix = {arXiv},
       eprint = {astro-ph/0201098},
 primaryClass = {astro-ph},
       adsurl = {https://ui.adsabs.harvard.edu/abs/2002Sci...295...82K}
}

@ARTICLE{kritos2024supe,
       author = {{Kritos}, Konstantinos and {Beckmann}, Ricarda S. and {Silk}, Joseph and {Berti}, Emanuele and {Yi}, Sophia and {Volonteri}, Marta and {Dubois}, Yohan and {Devriendt}, Julien},
        title = "{Supermassive black hole growth in hierarchically merging nuclear star clusters}",
      journal = {arXiv e-prints},
         year = 2024,
        month = dec,
          eid = {arXiv:2412.15334},
        pages = {arXiv:2412.15334},
          doi = {10.48550/arXiv.2412.15334},
archivePrefix = {arXiv},
       eprint = {2412.15334},
 primaryClass = {astro-ph.GA},
       adsurl = {https://ui.adsabs.harvard.edu/abs/2024arXiv241215334K}
}

@article{Juod_balis_2024,
   title={A dormant overmassive black hole in the early Universe},
   volume={636},
   ISSN={1476-4687},
   url={http://dx.doi.org/10.1038/s41586-024-08210-5},
   DOI={10.1038/s41586-024-08210-5},
   number={8043},
   journal={Nature},
   publisher={Springer Science and Business Media LLC},
   author={Juodžbalis, Ignas and Maiolino, Roberto and Baker, William M. and Tacchella, Sandro and Scholtz, Jan and D’Eugenio, Francesco and Witstok, Joris and Schneider, Raffaella and Trinca, Alessandro and Valiante, Rosa and DeCoursey, Christa and Curti, Mirko and Carniani, Stefano and Chevallard, Jacopo and de Graaff, Anna and Arribas, Santiago and Bennett, Jake S. and Bourne, Martin A. and Bunker, Andrew J. and Charlot, Stéphane and Jiang, Brian and Koudmani, Sophie and Perna, Michele and Robertson, Brant and Sijacki, Debora and Übler, Hannah and Williams, Christina C. and Willott, Chris},
   year={2024},
   month=dec, pages={594–597} }

@article{Maiolino_2024,
   title={JADES: The diverse population of infant black holes at 4 &lt; z &lt; 11: Merging, tiny, poor, but mighty},
   volume={691},
   ISSN={1432-0746},
   url={http://dx.doi.org/10.1051/0004-6361/202347640},
   DOI={10.1051/0004-6361/202347640},
   journal={Astronomy \& Astrophysics},
   publisher={EDP Sciences},
   author={Maiolino, Roberto and Scholtz, Jan and Curtis-Lake, Emma and Carniani, Stefano and Baker, William and de Graaff, Anna and Tacchella, Sandro and Übler, Hannah and D’Eugenio, Francesco and Witstok, Joris and Curti, Mirko and Arribas, Santiago and Bunker, Andrew J. and Charlot, Stéphane and Chevallard, Jacopo and Eisenstein, Daniel J. and Egami, Eiichi and Ji, Zhiyuan and Jones, Gareth C. and Lyu, Jianwei and Rawle, Tim and Robertson, Brant and Rujopakarn, Wiphu and Perna, Michele and Sun, Fengwu and Venturi, Giacomo and Williams, Christina C. and Willott, Chris},
   year={2024},
   month=nov, pages={A145} }

@ARTICLE{Marsh,
       author = {{Marsh}, David J.~E. and {Pop}, Ana-Roxana},
        title = "{Axion dark matter, solitons and the cusp-core problem}",
      journal = {\mnras},
         year = 2015,
        month = aug,
       volume = {451},
       number = {3},
        pages = {2479-2492},
          doi = {10.1093/mnras/stv1050},
archivePrefix = {arXiv},
       eprint = {1502.03456},
 primaryClass = {astro-ph.CO},
       adsurl = {https://ui.adsabs.harvard.edu/abs/2015MNRAS.451.2479M}
}

@ARTICLE{Bucciotti_2023,
       author = {{Bucciotti}, Bruno and {Trincherini}, Enrico},
        title = "{interplay between black holes and ultralight dark matter: analytic solutions}",
      journal = {Journal of High Energy Physics},
         year = 2023,
        month = nov,
       volume = {2023},
       number = {11},
          eid = {193},
        pages = {193},
          doi = {10.1007/JHEP11(2023)193},
archivePrefix = {arXiv},
       eprint = {2309.02482},
 primaryClass = {hep-th},
       adsurl = {https://ui.adsabs.harvard.edu/abs/2023JHEP...11..193B}
}

@ARTICLE{Maiolino2025,
       author = {{Maiolino}, Roberto and {Uebler}, Hannah and {D'Eugenio}, Francesco and {Scholtz}, Jan and {Juodzbalis}, Ignas and {Perna}, Michele and {Bromm}, Volker and {Dayal}, Pratika and {Koudmani}, Sophie and {Liu}, Boyuan and {Schneider}, Raffaella and {Sijacki}, Debora and {Valiante}, Rosa and {Trinca}, Alessandro and {Zhang}, Saiyang and {Volonteri}, Marta and {Inayoshi}, Kohei and {Carniani}, Stefano and {Nakajima}, Kimihiko and {Isobe}, Yuki and {Witstok}, Joris and {Jones}, Gareth C. and {Tacchella}, Sandro and {Arribas}, Santiago and {Bunker}, Andrew and {Cataldi}, Elisa and {Charlot}, Stephane and {Cresci}, Giovanni and {Curti}, Mirko and {Fabian}, Andrew C. and {Katz}, Harley and {Kumari}, Nimisha and {Laporte}, Nicolas and {Mazzolari}, Giovanni and {Robertson}, Brant and {Sun}, Fengwu and {Rodriguez Del Pino}, Bruno and {Venturi}, Giacomo},
        title = "{A black hole in a near-pristine galaxy 700 million years after the Big Bang}",
      journal = {arXiv e-prints},
         year = 2025,
        month = may,
          eid = {arXiv:2505.22567},
        pages = {arXiv:2505.22567},
archivePrefix = {arXiv},
       eprint = {2505.22567},
 primaryClass = {astro-ph.GA},
       adsurl = {https://ui.adsabs.harvard.edu/abs/2025arXiv250522567M}
}

@ARTICLE{Greene2024,
       author = {{Greene}, Jenny E. and {Labbe}, Ivo and {Goulding}, Andy D. and {Furtak}, Lukas J. and {Chemerynska}, Iryna and {Kokorev}, Vasily and {Dayal}, Pratika and {Volonteri}, Marta and {Williams}, Christina C. and {Wang}, Bingjie and {Setton}, David J. and {Burgasser}, Adam J. and {Bezanson}, Rachel and {Atek}, Hakim and {Brammer}, Gabriel and {Cutler}, Sam E. and {Feldmann}, Robert and {Fujimoto}, Seiji and {Glazebrook}, Karl and {de Graaff}, Anna and {Khullar}, Gourav and {Leja}, Joel and {Marchesini}, Danilo and {Maseda}, Michael V. and {Matthee}, Jorryt and {Miller}, Tim B. and {Naidu}, Rohan P. and {Nanayakkara}, Themiya and {Oesch}, Pascal A. and {Pan}, Richard and {Papovich}, Casey and {Price}, Sedona H. and {van Dokkum}, Pieter and {Weaver}, John R. and {Whitaker}, Katherine E. and {Zitrin}, Adi},
        title = "{UNCOVER Spectroscopy Confirms the Surprising Ubiquity of Active Galactic Nuclei in Red Sources at z > 5}",
      journal = {\apj},
         year = 2024,
        month = mar,
       volume = {964},
       number = {1},
          eid = {39},
        pages = {39},
          doi = {10.3847/1538-4357/ad1e5f},
archivePrefix = {arXiv},
       eprint = {2309.05714},
 primaryClass = {astro-ph.GA},
       adsurl = {https://ui.adsabs.harvard.edu/abs/2024ApJ...964...39G}
}

@ARTICLE{Matthee24,
       author = {{Matthee}, Jorryt and {Naidu}, Rohan P. and {Brammer}, Gabriel and {Chisholm}, John and {Eilers}, Anna-Christina and {Goulding}, Andy and {Greene}, Jenny and {Kashino}, Daichi and {Labbe}, Ivo and {Lilly}, Simon J. and {Mackenzie}, Ruari and {Oesch}, Pascal A. and {Weibel}, Andrea and {Wuyts}, Stijn and {Xiao}, Mengyuan and {Bordoloi}, Rongmon and {Bouwens}, Rychard and {van Dokkum}, Pieter and {Illingworth}, Garth and {Kramarenko}, Ivan and {Maseda}, Michael V. and {Mason}, Charlotte and {Meyer}, Romain A. and {Nelson}, Erica J. and {Reddy}, Naveen A. and {Shivaei}, Irene and {Simcoe}, Robert A. and {Yue}, Minghao},
        title = "{Little Red Dots: An Abundant Population of Faint Active Galactic Nuclei at z {\ensuremath{\sim}} 5 Revealed by the EIGER and FRESCO JWST Surveys}",
      journal = {ApJ},
         year = 2024,
        month = mar,
       volume = {963},
       number = {2},
          eid = {129},
        pages = {129},
          doi = {10.3847/1538-4357/ad2345},
archivePrefix = {arXiv},
       eprint = {2306.05448},
 primaryClass = {astro-ph.GA},
       adsurl = {https://ui.adsabs.harvard.edu/abs/2024ApJ...963..129M}
}

@ARTICLE{Maiolino24,
       author = {{Maiolino}, Roberto and {Scholtz}, Jan and {Curtis-Lake}, Emma and {Carniani}, Stefano and {Baker}, William and {de Graaff}, Anna and {Tacchella}, Sandro and {{\"U}bler}, Hannah and {D'Eugenio}, Francesco and {Witstok}, Joris and {Curti}, Mirko and {Arribas}, Santiago and {Bunker}, Andrew J. and {Charlot}, St{\'e}phane and {Chevallard}, Jacopo and {Eisenstein}, Daniel J. and {Egami}, Eiichi and {Ji}, Zhiyuan and {Jones}, Gareth C. and {Lyu}, Jianwei and {Rawle}, Tim and {Robertson}, Brant and {Rujopakarn}, Wiphu and {Perna}, Michele and {Sun}, Fengwu and {Venturi}, Giacomo and {Williams}, Christina C. and {Willott}, Chris},
        title = "{JADES: The diverse population of infant black holes at 4 < z < 11: Merging, tiny, poor, but mighty}",
      journal = {\aap},
         year = 2024,
        month = nov,
       volume = {691},
          eid = {A145},
        pages = {A145},
          doi = {10.1051/0004-6361/202347640},
archivePrefix = {arXiv},
       eprint = {2308.01230},
 primaryClass = {astro-ph.GA},
       adsurl = {https://ui.adsabs.harvard.edu/abs/2024A&A...691A.145M}
}

@article{Agarwal_2014,
   title={The First Billion Years project: birthplaces of direct collapse black holes},
   volume={443},
   ISSN={0035-8711},
   url={http://dx.doi.org/10.1093/mnras/stu1112},
   DOI={10.1093/mnras/stu1112},
   number={1},
   journal={Monthly Notices of the Royal Astronomical Society},
   publisher={Oxford University Press (OUP)},
   author={Agarwal, Bhaskar and Dalla Vecchia, Claudio and Johnson, Jarrett L. and Khochfar, Sadegh and Paardekooper, Jan-Pieter},
   year={2014},
   month=jul, pages={648–657} }

@ARTICLE{trinca2024,
       author = {{Trinca}, Alessandro and {Valiante}, Rosa and {Schneider}, Raffaella and {Juod{\v{z}}balis}, Ignas and {Maiolino}, Roberto and {Graziani}, Luca and {Lupi}, Alessandro and {Natarajan}, Priyamvada and {Volonteri}, Marta and {Zana}, Tommaso},
        title = "{Episodic super-Eddington accretion as a clue to Overmassive Black Holes in the early Universe}",
      journal = {arXiv e-prints},
         year = 2024,
        month = dec,
          eid = {arXiv:2412.14248},
        pages = {arXiv:2412.14248},
          doi = {10.48550/arXiv.2412.14248},
archivePrefix = {arXiv},
       eprint = {2412.14248},
 primaryClass = {astro-ph.GA},
       adsurl = {https://ui.adsabs.harvard.edu/abs/2024arXiv241214248T}
}

@article{piana2024,
    author = {Piana, Olmo and Pu, Hung-Yi and Wu, Kinwah},
    title = {Super-Eddington accretion in high-redshift black holes and the emergence of jetted AGN},
    journal = {Monthly Notices of the Royal Astronomical Society},
    volume = {530},
    number = {2},
    pages = {1732-1748},
    year = {2024},
    month = {03},
    issn = {0035-8711},
    doi = {10.1093/mnras/stae851},
    url = {https://doi.org/10.1093/mnras/stae851},
    eprint = {https://academic.oup.com/mnras/article-pdf/530/2/1732/57273315/stae851.pdf},
}

@article{spitzer1969,
  title={Equipartition and the formation of compact nuclei in spherical stellar systems},
  author={Spitzer Jr, Lyman},
  journal={Astrophysical Journal, vol. 158, p. L139},
  volume={158},
  pages={L139},
  year={1969}
}

@article{Hopkins_2018,
   title={FIRE-2 simulations: physics versus numerics in galaxy formation},
   volume={480},
   ISSN={1365-2966},
   url={http://dx.doi.org/10.1093/mnras/sty1690},
   DOI={10.1093/mnras/sty1690},
   number={1},
   journal={Monthly Notices of the Royal Astronomical Society},
   publisher={Oxford University Press (OUP)},
   author={Hopkins, Philip F and Wetzel, Andrew and Kereš, Dušan and Faucher-Giguère, Claude-André and Quataert, Eliot and Boylan-Kolchin, Michael and Murray, Norman and Hayward, Christopher C and Garrison-Kimmel, Shea and Hummels, Cameron and Feldmann, Robert and Torrey, Paul and Ma, Xiangcheng and Anglés-Alcázar, Daniel and Su, Kung-Yi and Orr, Matthew and Schmitz, Denise and Escala, Ivanna and Sanderson, Robyn and Grudić, Michael Y and Hafen, Zachary and Kim, Ji-Hoon and Fitts, Alex and Bullock, James S and Wheeler, Coral and Chan, T K and Elbert, Oliver D and Narayanan, Desika},
   year={2018},
   month=June, pages={800–863} }

@article{Pillepich_2017,
   title={Simulating galaxy formation with the IllustrisTNG model},
   volume={473},
   ISSN={1365-2966},
   url={http://dx.doi.org/10.1093/mnras/stx2656},
   DOI={10.1093/mnras/stx2656},
   number={3},
   journal={Monthly Notices of the Royal Astronomical Society},
   publisher={Oxford University Press (OUP)},
   author={Pillepich, Annalisa and Springel, Volker and Nelson, Dylan and Genel, Shy and Naiman, Jill and Pakmor, Rüdiger and Hernquist, Lars and Torrey, Paul and Vogelsberger, Mark and Weinberger, Rainer and Marinacci, Federico},
   year={2017},
   month=Oct, pages={4077–4106} }

@article{Weinberger_2018,
   title={Supermassive black holes and their feedback effects in the IllustrisTNG simulation},
   volume={479},
   ISSN={1365-2966},
   url={http://dx.doi.org/10.1093/mnras/sty1733},
   DOI={10.1093/mnras/sty1733},
   number={3},
   journal={Monthly Notices of the Royal Astronomical Society},
   publisher={Oxford University Press (OUP)},
   author={Weinberger, Rainer and Springel, Volker and Pakmor, Rüdiger and Nelson, Dylan and Genel, Shy and Pillepich, Annalisa and Vogelsberger, Mark and Marinacci, Federico and Naiman, Jill and Torrey, Paul and Hernquist, Lars},
   year={2018},
   month=June, pages={4056–4072} }

@article{Angl_s_Alc_zar_2017,
   title={Black holes on FIRE: stellar feedback limits early feeding of galactic nuclei},
   volume={472},
   ISSN={1745-3933},
   url={http://dx.doi.org/10.1093/mnrasl/slx161},
   DOI={10.1093/mnrasl/slx161},
   number={1},
   journal={Monthly Notices of the Royal Astronomical Society: Letters},
   publisher={Oxford University Press (OUP)},
   author={Anglés-Alcázar, Daniel and Faucher-Giguère, Claude-André and Quataert, Eliot and Hopkins, Philip F. and Feldmann, Robert and Torrey, Paul and Wetzel, Andrew and Kereš, Dušan},
   year={2017},
   month=Oct, pages={L109–L114} }

@article{atmabacak_2022,
   title={Black hole–galaxy scaling relations in FIRE: the importance of black hole location and mergers},
   volume={511},
   ISSN={1365-2966},
   url={http://dx.doi.org/10.1093/mnras/stac040},
   DOI={10.1093/mnras/stac040},
   number={1},
   journal={Monthly Notices of the Royal Astronomical Society},
   publisher={Oxford University Press (OUP)},
   author={Çatmabacak, Onur and Feldmann, Robert and Anglés-Alcázar, Daniel and Faucher-Giguère, Claude-André and Hopkins, Philip F and Kereš, Dušan},
   year={2022},
   month=Jan, pages={506–535} }

@article{Cochrane2023,
  author = {Cochrane, R. K. and Hayward, C. C. and Angl{\'e}s-Alc{\'a}zar, D. and Somerville, R. S.},
  title = {Predicting sub-millimetre flux densities from global galaxy properties},
  journal = {Monthly Notices of the Royal Astronomical Society},
  volume = {518},
  number = {4},
  pages = {5522--5539},
  year = {2023},
  doi = {10.1093/mnras/stac3472},
  eprint = {2211.11702},
  archivePrefix = {arXiv},
  primaryClass = {astro-ph.GA}
}

@ARTICLE{Power2003,
       author = {{Power}, C. and {Navarro}, J.~F. and {Jenkins}, A. and {Frenk}, C.~S. and {White}, S.~D.~M. and {Springel}, V. and {Stadel}, J. and {Quinn}, T.},
        title = "{The inner structure of {\ensuremath{\Lambda}}CDM haloes - I. A numerical convergence study}",
      journal = {\mnras},
         year = 2003,
        month = jan,
       volume = {338},
       number = {1},
        pages = {14-34},
          doi = {10.1046/j.1365-8711.2003.05925.x},
archivePrefix = {arXiv},
       eprint = {astro-ph/0201544},
 primaryClass = {astro-ph},
       adsurl = {https://ui.adsabs.harvard.edu/abs/2003MNRAS.338...14P}
}
\bibliographystyle{aasjournal} 

\end{document}